\documentclass[11pt]{article}
\usepackage{fullpage}
\usepackage{framed}
\usepackage{mdframed}

\usepackage{amsmath}
\usepackage{amssymb}

\usepackage[utf8]{inputenc} % allow utf-8 input
\usepackage[T1]{fontenc}    % use 8-bit T1 fonts
\usepackage{hyperref}       % hyperlinks
\usepackage{url}            % simple URL typesetting
\usepackage{booktabs}       % professional-quality tables
\usepackage{amsfonts}       % blackboard math symbols
\usepackage{nicefrac}       % compact symbols for 1/2, etc.
\usepackage{microtype}      % microtypography

\def\defeq{\overset{\Delta}{=}}  % Equal with triangle
\def\cl{\mathsf{cl\ }}  % Closure
\def\d{\mathsf{d}}  % Differential operator
\def\D{\mathsf{D}}  % Derivative operator

\def\H{\mathcal{H}}  % Hilbert space
\def\E{\mathbb{E}}  % Expectation
\def\P{\mathbb{P}}  % Probability Measure
\def\R{\mathbb{R}}  % Set of real numbers
\def\N{\mathbb{N}}  % Set of natural numbers
\def\T{\mathsf{T}}  % Transpose notation
\def\ind{\mathbf{1}}  % Ones vector or indicator
\def\<{\langle}  % < Inner product
\def\>{\rangle}  % > Inner product

\def\B{\mathcal{B}}  % Contract set B_j
\def\A{\mathcal{A}}  % Targetting set A_i
\def\H{\mathcal{H}}  % Hamiltonian
\def\Tt{\mathcal{T}_t^N}  % Set of active time indices
\def\Ttj{\mathcal{T}_t^M}  % Set of active item types
\def\Tk{\mathcal{T}_k^N}  % Discrete analog of Tt
\def\L{\mathcal{L}}  % Set of active time indices
\def\mS{\mathcal{S}}  % A set S
\def\mC{\mathcal{C}}  % A Contract /C/ = (T, C, S)
\def\mR{\mathcal{R}}  % A collection of sets /R/ = {R_j}
\def\mA{\mathcal{A}}  % A_i
\def\mB{\mathcal{B}}  % B_j
\def\Ws{\overline{W}_{\sigma}}  % Smoothed W
\def\ffs{\overline{f}_{\sigma}}  % Smoothed f
\def\bM{\mathbf{M}}  % The whole collection of bidders
\def\bW{\mathbf{W}}  % The win probability
\def\Ljk{\overline{\Lambda}_{jk}}
\def\Wjk{\overline{W}_{jk}}
\def\fjk{\overline{f}_{jk}}

\newcommand{\RNum}[1]{\uppercase\expandafter{\romannumeral #1\relax}}  % Roman numerals

\usepackage[backend=biber,sorting=nyt,firstinits=true,maxnames=10000]{biblatex}
\bibliography{refs}

\newtheorem{thm}{{\bf Theorem}}[section]

\newtheorem{prop}{{\bf Proposition}}[section]

\newtheorem{remark}{{\bf Remark}}[section]
\newtheorem{assumption}{{\bf Assumption}}[section]
\newtheorem{tdef}{{\bf Definition}}[section]
\newtheorem{example}{{\bf Example}}[section]

\usepackage{enumitem}
\usepackage{etex,etoolbox}
\usepackage{hyperref}
\usepackage{framed}

\usepackage{booktabs}
\usepackage{thmtools}
\usepackage{environ}

\usepackage{graphicx}
\usepackage{caption}
\usepackage{subcaption}

\usepackage[linesnumbered, vlined, plain]{algorithm2e}

\title{Optimal Real-time Bidding Policies for Contract Fulfillment in Second Price Auctions}

\author{
  R. J. Kinnear\thanks{Relevant code to be made available at github.com/RJTK} \\
  \small Department of Electrical and Computer Engineering\\
  \small University of Waterloo\\
  \small Waterloo, ON, N2L 3G1 \\
  \small \texttt{ryan@kinnear.ca} \and
  R. R. Mazumdar \\
  \small Department of Electrical and Computer Engineering\\
  \small University of Waterloo\\
  \small Waterloo, ON, N2L 3G1 \\
  \small \texttt{mazum@uwaterloo.ca} \and
  P. Marbach \\
  \small Department of Computer Science\\
  \small University of Toronto\\
  \small Toronto, ON, M5S 3G4\\
  \small \texttt{marbach@cs.toronto.edu}
}

\begin{document}
\maketitle

\begin{abstract}
  We study a real-time bidding problem resulting from a set of
  contractual obligations stipulating that a firm win a specified
  number of heterogenous impressions or ad placements over a defined duration in a realtime
  auction.  The contracts specify item targeting criteria (which may
  be overlapping), and a supply requirement. 
  Using the Pontryagin maximum principle, we show that the resulting
  continuous time and time inhomogenous planning problem can be
  reduced into a finite dimensional convex optimization problem and
  solved to optimality.  In addition, we provide algorithms to update
  the bidding plan over time via a receding horizon.  Finally, we
  provide numerical results based on real data and show a connection
  to production-transportation problems.
\end{abstract}

% keywords can be removed
\paragraph{keywords}
  Computational Advertising; Realtime Bidding; Optimal
  Control; Auction Theory; Second Price Auction; Production
  Transportation Problem
  
\paragraph{Acknowledgement}
We acknowledge the support of the Natural Sciences and
  Engineering Research Council of Canada (NSERC), [funding reference
  number 518418-2018].  Cette recherche a été financée par le Conseil
  de recherches en sciences naturelles et en génie du Canada (CRSNG),
  [numéro de référence 518418-2018].
  
\newpage
\section{Introduction}
Online advertising constitutes a significant part of today's
advertising landscape.  The total amount of money spent directly on
internet advertising (the largest advertising segment, far surpassing
competitors like TV and print) in 2018, according to The Interactive
Advertising Bureau \cite{iab2018iab} exceeded \$100b, and display
advertising alone (as opposed to sponsored search) accounted for
roughly 40\% of this total.  Moreover, year on year growth rates
remain extremely high: revenue in 2018 grew by 21.8\% over 2017, and
compounded annual growth rate between 2012-2018 exceeds 45\%.

This advertising market consists primarily of two groups of players:
publishers (e.g., website operators, apps) and advertisers.  The goal
of the advertisers in this setting is to get their messaging in front
of the visitors to publisher websites or app users, generally for the
purposes of generating either brand awareness in the long-term or
immediate purchasing decisions in the short-term
\cite{choi2020online}.  Matching ads to users is facilitated by
\textit{ad exchanges} like Google AdX \cite{mansour2012doubleclick},
which allow publishers seeking to sell space on their website to
solicit requests from advertisers.  The exchanges implement a clearing
mechanism in the form of an auction.

This paper focuses on a class of on-line advertising problems known as
real-time bidding (RTB) auctions ~\cite{choi2020online}.  RTB
constitutes about 35\% of the entire display advertising market, with
the remainder of the market consisting of traditional fixed
advertising contracts which require a publisher to display an
advertiser's content for a pre-negotiated period of time
\cite{chen2014dynamic}.  RTB is characterized by auctions which occur
every time a user visits a web page or opens an app, the ``item'' up
for sale in the auction is an ad space that the winner of the auction
fills with their advertising content.  These items are also referred
to as ``impressions''.  Bidders that lose the auction for a particular
item need to wait for the next opportunity.  The entire process
beginning with the arrival of a user, to the decision about the winner
of the auction and the display of the winner's content, takes place in
around 100ms.

Every auction taking place in RTB is sealed bid, and each bidder
submits only a single bid.  Moreover, bidding data is censored:
bidders are not informed about the bid that won the auction, unless
they are themselves the winner.  For the winning bidder (which is
always the highest bidder) their actual payment depends on the type of
auction the exchange is running.  The two most prevalent
\textit{basic} auction types are {\em first price} auctions wherein
the winner pays what they bid, and {\em second price}, also known as
{\em Vickrey auctions} \cite{vickrey1961counterspeculation}, wherein
the winner pays the second highest bid.  The true mechanism in
practice often incorporates both types with a first price auction
below a (``soft'') floor \cite{zeithammer2019soft} and a second price
auction otherwise.

In practice, advertisers use the services of aggregators called Demand
Side Platforms (DSPs) (see e.g. \cite{wang2017display} for additional
information) which participate in RTB on the advertiser's behalf.  It
is common practice for a DSP to enter a contract with an advertiser
which stipulates an up-front fee be paid for the guarantee that a
minimum number of ads be displayed to targeted segments of the
population (age, sex, location, other preferences, etc.).  The
segments are referred to as the targeting criteria and each such
contract is referred to as a campaign.  From the perspective of the
DSP, the optimal bids should minimize the total cost to obtain the
required number of impressions.

The problem that we address in this paper are optimal (i.e., cost
minimizing) bidding strategies for DSPs to fulfill their contracts.
Typically, DSPs handle hundreds of campaigns simultaneously and the
targeting criteria of the campaigns may overlap.  This induces a
problem where on every bid request the DSP receives from the ad
exchange, they must decide, based on the characteristics of the item,
what price to bid and which contract the impression, if it is won, it
should be allocated towards fulfilling.  An important characteristic
of our perspective is that it does not involve item valuations: the
DSPs we consider seeks to fulfill acquisition contracts, not to
maximize their valuation of items won.  Once a contract has been
agreed to, the DSP \textit{must} fulfill it's obligations.

\subsection{Literature Review}
Problems of optimal bidding have been addressed at different levels of
generality and from various perspectives.  Early works addressed the
problem in the context of a single campaign and budget constraints
\cite{gummadi2013optimal} on an infinite horizon.
They assumed that the prices of impressions arrive as an i.i.d process
from an unknown distribution and the goal was to maximize the utility
(or valuation) subject to constraints on the budget for the ergodic
and discounted cost criteria.  The optimal strategy (which is
stationary in this case) is to bid according to a \textit{shaded}
(i.e. reduced) item valuation where the shading factor arises from a
Lagrange multiplier associated with the budget constraint and depends
on the unknown distribution of the prices.  In
\cite{jiang2014bidding}, they address the problem of determining the
optimal shading factor by using a stochastic approximation algorithm
in an i.i.d. price setting.  The case when there are a large number of
bidders was studied in \cite{balseiro2015repeated} where a mean-field
approach based on independence of the bidders was assumed.  The optimal
structure of the bids is similar to \cite{gummadi2013optimal}.

In ~\cite{zhang2014optimal} they consider a problem with many
campaigns but with identical targeting criteria with the goal being to
maximize the number of impressions subject to budget and risk
constraints where risk is taken as a variance constraint on the total
number of items.  In \cite{zhang2012joint} they consider the problem
with multiple campaigns with non-overlapping targeting criteria in the
distinct but related context of sponsored search.

%%% This is an inaccurate characterization of the paper, presumably a
%%% typo on the citation
% In \cite{balseiro2019learning} they consider a many campaigns case with
% overlapping targeting criteria where they propose heuristics for
% bidding and allocating bid requests to the various campaigns subject
% to a constraint on the budget.

Many other algorithms have been been deployed on the problem of
optimal bidding, including classical feedback control in the work of
\cite{zhang2016feedback, karlsson2014adaptive} where they seek to
track certain keep performance indicators and \cite{cai2017real,
  grislain2019recurrent, wu2018budget} which utilize the Markov
Decision framework.

In the recent work of \cite{marbach_bidding_2020}, the optimal
contract management problem with multiple campaigns and overlapping
targeting criteria was studied for the static case of optimizing over
one duration. This is equivalent to apportioning equal impression
requirements to each duration in the term of a contract.  One of the
key insights was the need for a \textit{supply curve} for each
targeting criterion (see also \cite{karlsson2016control}).  The supply
curve is simply a right continuous increasing function indicating the
average number of impressions (or estimate thereof) that will be won
given a particular bid. We discuss this issue later.

An important attendant problem is that of estimating and forecasting
market prices, referred to as ``bid landscape
forecasting''~\cite{cui2011bid, wu2015predicting, zhang2016bid,
  wang2016functional, ghosh2020scalable} which ultimately falls into
the domain of statistics and machine learning applications.  This 
 important problem is not the focus of this paper.

\subsection{Contributions}
We formulate the contract fulfillment problem faced by a DSP as a
continuous time optimal control problem and provide algorithms
specifying how to bid on any given bid request.  We show that the most
general case can be reduced to solving the time homogeneous problem
and that this solution can be computed through the solution of a
convex optimization problem.  This is a direct generalization of
\cite{marbach_bidding_2020} to account for differing contract
deadlines and time-inhomogenous supply curves.  Moreover, we also show
that the problem of \cite{marbach_bidding_2020} can be addressed via
convex optimization.

We show that the time dependent problem addressed in this paper can be
solved via the application of optimal control theory, which allows for
bids at earlier periods to be appropriately readjusted to account for
future anticipated changes in supply or price.  In order to account
for the moment-to-moment adaptation, our solution may serve as a set
point for the classical regulators studied for example by
\cite{karlsson2014adaptive, zhang2016feedback}.  Moreover, while our
basic formulation doesn't endogenously account for stochasticity in
the environment, this is accounted for via a receding horizon
\cite{cannon2016model}.

Our formulation can also be seen as analogous to a continuous time
Production-Transportation problem \cite{leblanc1974transportation,
  sharp1970decomposition} with transportation costs taking values of
either $0$ or $\infty$.

\subsection{Outline}
We begin by discussing a simple market model and how the idea of a
\textit{supply curve} (Section
\ref{sec:market_model_and_supply_curves}) naturally arises and serves
essentially as an {\em information} state for the problem.  In Section
\ref{sec:cost_functions} we introduce the relevant cost functions and
auction mechanism. We focus is on second price auctions 
though many of our results can apply to some more
general auctions that will be explored elsewhere.

Section \ref{sec:time_constrained_impression_contracts} formally
introduces our problem and Section
\ref{sec:target_partitioning} discusses segmenting the market
according to the needs of the contract management problem.

Section \ref{sec:optimal_control} examines a simple special case and
illustrates the receding horizon method through an analytically
tractable example.  Section \ref{sec:general_case} provides a concrete
formulation as a continuous time optimal control problem, and
rigorously establishes the existence and optimality of solutions to
our problem along with the basic properties which enable the
transformation into a finite convex optimization problem.

Our final result in Theorem \ref{thm:maintheorem} is  that
the entire continuous time portfolio management problem can be solved
to optimality via a finite convex optimization problem.

In Section \ref{sec:simulations} we illustrate our methods through
application on publicly available IPinYou dataset
\cite{liao2014ipinyou, zhang2014real}, and compare against a solution
obtained without direct consideration of time inhomogeneity. The cost
savings by considering the dynamics is about $10\%$.

\section{Market Model and Supply Curves}
\label{sec:market_model_and_supply_curves}
In this section we outline a simple market model that will lead to an
understanding of the properties of what we will call the
\textit{supply curve} $W_j(x, t)$, indicating the \textit{average}
instantaneous (at time $t$) rate of items accumulated by bidding $x$
on every item of type $j$.  This function will describe what is
analogous to an {\em information state} for a bidder participating in
the market, i.e., $W_j(x, t)$ encodes all of the
information necessary for them to make informed decisions about
bidding.  The function $W_j(x, t)$ naturally arises from the actions
taken by bidders who are present in the market at time $t$.

Since empirical data \cite{yuan2013real, zhang2014real} demonstrates
clear cyclic and time varying behaviour in sale prices and item
volumes, we explicitly consider time dependence in $W_j(x, t)$ to
capture these dynamics and allow bidders to \textit{plan} for the
future states of the market.

\subsection{Market Model}
\label{sec:market_model}
Suppose that we have a generic real-time auction exchange dealing in
heterogeneous items of types\footnote{Throughout, we use the notation
  $[N] = \{1, \ldots, N\}$} $j \in [M]$.  It may be that separate type
$j$ items are still distinguishable by participants in reality, but we
do not model any intra-type distinctions.  At any given time, there is
a large group of bidders (or ``participants'') who participate in the
auction exchange, this group of bidders is subject to change over
time.  The items arrive (one by one) over time to the exchange, with
each arrival triggering a bid to be submit by some (but not
necessarily all) of the currently present participants.

Suppose that at fixed time $t$ there are $N(t)$ bidders
($i = 1, 2, \ldots, N(t)$) present at the auction exchange, and that
for any bidder $i$, their behaviour is described by $M$ (bid, rate)
tuples: $\{(b_{ij}, r_{ij})\}_{j = 1}^M$.  The quantity
$r_{ij} \in (0, 1)$ indicates that if an item of type $j$ arrives,
bidder $i$ will bid on it with probability $r_{ij}$, independently of
all other bidders; possible interpretations being that items of type
$j$ match $i$'s interests with probability $r_{ij}$, or that they only
bid on a fraction of each item type in order to spread out their
budget over time.  The bid placed by $i$ for items of type $j$ is
given by $b_{ij} \in \R_+$.  The entire state of the market at time
$t$ including $N(t)$ and $b, r$ will be denoted $\bM_t$.

Since the winner of the auction is the individual submitting the
largest bid, we denote by $\bW_j(t)$ the \textit{price process} which
at time $t$ takes the value of the largest bid that would be submit by
the participants present in the auction exchange if at time $t$ an
item arrived to be bid upon.  The probability that an exogenous bidder
would win an item of type $j$ arriving at the instant $t$ if they
placed the bid $x$ is therefore a cumulative distribution function
$\P\{W_j(t) \le x\}$.  We can determine the exact form of this
c.d.f. when the market state is fixed as follows:

\begin{prop}[Win Probability Properties]
  \label{prop:supply_curve_properties}
  If at the fixed time $t$ the market $\bM_t$ consists of $N$
  participants, namely $\{(r_i, b_i)\}_{i = 1}^N$, and suppose each $b_i$
  is distinct, then the probability of an exogenous participant
  (denoted by $0$) winning an item with the bid $x$ and where ties are
  always settled in $0$'s favour, is a cumulative distribution
  function denoted by
  $W_j^\bM(x, t) \defeq \P\{\bW_j(t) \le x\ |\ \bM_t \}$.  In the
  model outline above, this function is given explicitly by

  \begin{equation}
    \label{eqn:fixed_N_W}
    W^\bM_j(x, t) = \textup{exp}\Big(-\sum_{i = 1}^N \phi(r_{ij})\ind_{(x, \infty)}(b_{ij})\Big)\ind_{\R_+}(x),
  \end{equation}

  where $\phi(r) \defeq -\ln (1 - r)$.
\end{prop}

The proof can be found in the Appendix, see \ref{sec:proofs}.

\subsection{Demand Side}
\label{sec:market_dynamics}
The number of participants $N(t)$ at any time is determined from the
stochastic dynamics of how participants arrive and the time spent in
the bidding process. Suppose participants arrive as a Poisson process
of rate $\rho$ and let us assume that the amount of time in the bidding
process is of unit duration (100ms in reality). Then, in equilibrium,
the number of participants is given by the distribution of an
$M/G/\infty$ model that is a Poisson distribution with parameter
$\rho$ denoted by $\text{Po}(\rho)$, see for example
\cite{adan2002queueing}.

Therefore, suppose that at an item arrival
instant $t$, we have $N(t) \sim \text{Po}(\rho)$ bidders, where the parameter
$\rho > 0$.  Moreover, suppose that the (bid, rate) parameters of
participants are drawn independently from distributions $F_{B_j}$ and
$F_{R_j}$ respectively, that is, at any $t$, the parameters
of the $N(t)$ participants are specified by
$b_{ij} \overset{\text{i.i.d.}}{\sim} F_{B_j}, r_{ij}
\overset{\text{i.i.d.}}{\sim} F_{R_j}$.  This is sufficient to derive and
motivate a supply curve $W_j(x, t)$.

We consider the \textit{average} win probability
$W^\bM_j(x) \defeq \E[W_j^\bM(x, t)]$ that can be characterized as
follows where we suppress the index $j$.

\begin{prop}[Expected Win Rate]
  \label{prop:expected_win_rate}
  
  If the bid distribution $F_B$ admits a probability density, and
  $N(t) \sim \text{Po}(\rho)$, then we have

  \begin{equation}
    W^{ss}(x) \defeq \E[W^\bM(x, t)] = e^{-\rho_B(x) \E[r]}
  \end{equation}

  where $\rho_B(x) \defeq \rho (1 - F_B(x))$ and
  $\E[r] = \int_0^\infty(1 - F_R(x))dx$.  Hence, $W^{ss}(x)$ is a
  cumulative distribution function.
  \end{prop}
  
{\noindent \bf Proof:} See Appendix,  see \ref{sec:proofs}.

\subsection{Supply Side}
\label{sec:acquisition_functions}
So far, we have focused on the demand side (i.e., the bidders) of the
auction, leading to the win probability (for type $j$) function
$W_j^\bM(x, t)$ and it's average $W_j^{ss}(x)$ interpreted as the mean
in steady state.

Turning attention now to the supply side (i.e. arrival of items), let
us consider an arrival point process $A_j(t)$ (independent of
$\bM_t$) with a time dependent intensity $\lambda_j(t)$.  In the case
of internet advertising, where item arrivals correspond to users
visiting a web page, it is natural for the arrival rate to be time
dependent, and can naturally be expected to exhibit daily and weekly
cycles.

In this model then, if we are given a deterministic function
$x: [0, T] \rightarrow \R_+$, the expected number of wins for an agent
bidding according to $x$ can be calculated simply via Campbell's
formula (see \cite{mazumdar2009performance} for example)

\begin{equation}
  \E \int_0^T \ind_{\{\bW_j(t) \le x(t)\}} \d A_j(t) = \int_0^T \lambda_j(t) W_j^{ss}(x(t)) \d t.
\end{equation}

% \begin{remark}
%   The upshot of Campbell's formula is that for a random process $X(t)$
%   and an independent Poisson process $A(t)$ with intensity
%   $\lambda(t)$, if points of $A(t)$ occur at the random times
%   $t_1, t_2, \ldots$ then we have
%   $\E[X(t_1) + X(t_2) + \cdots] \defeq \E \int_0^T X(t) dA(t) =
%   \int_0^T \lambda(t) \E[X(t)]dt$. (see
%   e.g. \cite{mazumdar2009performance})
% \end{remark}

It is in this sense that $W_j(x, t) = \lambda_j(t) W_j^{ss}(x)$ is the
average number of items won instantaneously at time $t$ given a bid of
$x(t)$.  A time varying average win probability function $W_j(x, t)$
thus arises naturally in RTB, and integrating this function results
naturally in the \textit{average} number of items won with the fixed
bid path $x(t)$.  We will summarize these ideas later in Definition
\ref{def:supply_curve}.

\subsection{Cost Functions and Auctions}
\label{sec:cost_functions}
Before formulating our main problem (Section \ref{sec:general_case})
we need to define a cost function.  These functions arise most
naturally from the rules of the auction, which will always be
sealed-bid second price auctions wherein the item is sold immediately
after a single round of bidding.  Extensions to more general cost
functions is possible but subtle and will be revisited in future work.

We will denote by $f(x, t)$ the (estimate of the) \textit{expected}
cost of bidding $x$ on an item arriving at time $t$.  To explain the
second price auction mechanism suppose that the bids among $N$
participants are denoted $b_1, ..., b_N$.  Bidder $i$ will win the
auction with bid $b_i$ if $b_i \ge 0$ and
$b_i > \underset{j \ne i}{\text{max}}\ b_j,$ where we can break ties
randomly.  If $i$ is the winner, they pay
$\underset{j \ne i}{\text{max}}\ b_j,$ which is in general
\textit{less} than their own bid.

From here we can see that if for a particular bidder, the maximum of
competing bids is given by the random variable $Y \sim F_Y$, and they bid
the value $x \ge 0$, their expected payment is
\begin{equation*}
  \E\bigl[Y \ind[0 < Y < x]\bigr] = \int_0^x u \d F_Y(u).
\end{equation*}

In our context, the distribution of ``$Y$'' at time $t$ for items of
type $j$ is given by $W_j(x, t)$, modulo the supply rate normalization
in $W_j$.  Therefore, the expected cost of bidding $x$ on items of
type $j$ is, instantaneously at time $t$,
\begin{equation}
  \label{eqn:sp_cost}
  f_j(x, t) = \ind_{\R_+}(x) \int_0^x u \d W_j(u, t).
\end{equation}

We include $\ind_{\R_+}(x)$ since we will allow the domain of $x$ to
be all of $\R$.

% Do I even need this?
% For later use, we additionally point out the simple form of the cost
% function derivative, which follows directly from the fundamental
% theorem of calculus:

% \begin{remark}[Cost Derivative]
%   \label{rem:cost_derivative}
%   It is clear that if $W$ is differentiable with derivative $W'$ then
%   the cost function $f$ has derivative
%   \begin{equation}
%     \label{eqn:sp_derivative}
%     f'(x, t) = x W'(x, t),
%   \end{equation}

%   which implies that $f$ is, along with $W$, monotone.
% \end{remark}

\subsection{Randomized Bidding}
\label{sec:smooth_supply_curves}
Since there is a finite number of bidders participating, the nature of
the auction mechanism makes it very natural for $W_j(x, t)$ to exhibit
discontinuous \textit{jumps} (w.r.t. $x$).  Such discontinuities are
observed in real data, see e.g., \cite{karlsson2016control,
  grislain2019recurrent, marbach_bidding_2020}).  Discontinuities may
arise even in estimated supply curves e.g. if it is desirable to
estimate directly the location of jumps in market prices, or if the
estimates of $W_j$ are carried out via an histogram, which is
naturally discontinuous.  However, for the purposes of deriving
bidding strategies, it is desirable to work with continuous supply
curves.  To this end we will establish a means of implementing smooth
approximations to discontinuous supply curves via randomized bidding.
See also \cite{grislain2019recurrent, karlsson2016control} for earlier
applications of this idea.  An alternative approach is given in
\cite{marbach_bidding_2020} wherein the authors work more directly
with the discontinuous supply curves and establish a different type of
randomization scheme which doesn't attempt to smooth out the entire
curve.

In practice, randomization has the additional benefit of ``hedging''
against incorrectly estimating the locations of important jump
discontinuities, as well as providing a parameter (the amount of bid
noise) to probe the exploration-exploitation frontier if supply curve
estimation is to take place simultaneously with bidding.

By choosing a parameter $\sigma^2 > 0$, define the function (suppressing
the subscript $j$) $\Ws(x, t) = \E W(x + \sigma\mathcal{X}, t)$, where
$\mathcal{X} \sim \mathcal{N}(0, 1)$.  A DSP can implement the function
$\Ws$, which is a $\mathcal{C}_\infty$ function w.r.t. $x$ (this follows
directly from Leibniz's integral formula), by using randomized bids:
instead of placing the nominal bid $x$, sample a $\mathcal{N}(0, 1)$
variable $\mathcal{X}$ and then place the bid
$x + \sigma \mathcal{X}$.  This approximation has the secondary benefit of
ensuring that $\Ws$ is \textit{strictly} monotone increasing (hence
invertible).  We point out that we will generally have
$\Ws(x, t) \ne W(x, t)$, but that as long as $\sigma$ is small, the
difference is slight.  We formalize these notions in the following
proposition, a complete proof is relegated to the Appendix.

\begin{prop}[Smooth and Monotone Supply Curve]
  \label{prop:smoothed_probabilities}
  
  Let $W(x, t), t\geq 0$ be 
  $L$-Lipschitz in $x$ at all but at most $n$ points ($n$ that does  not depend on  $t$), and that
  $\underset{x \ge 0}{\text{sup}}\; W(x, t) = B(t) < \infty.$

  Then, for any $\epsilon > 0$ and any compact set $I \subset \R$,
  there exists $\sigma^2 > 0$ such that
  $\int_I |\Ws(x, t) - W(x, t)| \d x < \epsilon$ and if $W$ does not
  contain any jumps, such that
  $||\Ws(\cdot, t) - W(\cdot, t)||_\infty < \epsilon$.

  Moreover, $\Ws$ is a $\mathcal{C}_\infty$, strictly
  monotone increasing function and $x\Ws(x) \rightarrow 0$ as
  $x \rightarrow -\infty$.  In particular,
  $\cl \mathsf{range}\ \Ws(\cdot, t) = [0, B(t)]$.
\end{prop}

\noindent{\bf Proof}: See the Appendix,  see \ref{sec:proofs}.

The effect of randomization should also be accounted for in the cost
function.  Formally, the true average cost is given by
$\ffs(x) \defeq \E f(x + \sigma\mathcal{X})$.  However, for small
$\sigma$, we can make analogous statements as in Proposition
\ref{prop:smoothed_probabilities}, i.e., that
``$\ffs(x) \approx \ind_{\R_+}(x) \int_0^x u\Ws'(u)\d u$'', which justifies the use of
$\Ws$ as if it were the \textit{true} supply curve for a second price
auction.

Henceforth, we will posit existence of estimated supply curves and
assume that they are smooth and strictly monotone, keeping in mind
that these properties can be obtained from much less well behaved
curves through randomized bidding.  We can summarize the previous
notions as follows:

\begin{tdef}[Bid Path, Supply Curve, Cost Curve]
  \label{def:supply_curve}
  Let $x(t)$ denote the bid at time t. We refer to the sample-path
  $\{x(t)\}_{t\in[0,T]}$ as the \textit{bid path} where $[0,T]$
  denotes the duration of the contract.The bid path $x(.)$ thus
  represents a bidding policy.

  For a particular item of type $j$, the \textit{supply curve}
  $W_j(x, t)$ is the function such that for a fixed bidding path
  $x(t)$, the expected number of items won over the period $[0, T]$ is

  $$\int_0^T W_j(x(t), t) \d t.$$

  We assume that the range of $W_j$ satisfies
  $\cl \mathsf{range}\ W_j(\cdot, t) = [0, B_j(t)]$ for some
  $B_j(t) < \infty$, and for every $t\geq 0$,  $W(x,t)$ is strictly
  monotone increasing and twice differentiable in $x$ (recall
  Proposition \ref{prop:smoothed_probabilities}).  The derivative of
  the function $x \mapsto W_j(x, t)$ will be denoted $W_j'(x, t)$, as
  we have no need to refer to derivatives w.r.t. $t$.  Finally,
  $W_j(x, t) \ge 0$ and
  $xW_j(x, t) \rightarrow 0 \text{ as } x \rightarrow -\infty$.

  The function $f(x, t)$ is the average cost of bidding $x$
  at time $t$ and satisfies $f(x, t) = 0$ for any
  $x \le 0$.  It is continuously differentiable and strictly monotone
  for $x \ge 0$.
\end{tdef}

\subsection{The Cost of Acquisition}
\label{sec:production_cost}
Having defined a cost function $f_j(x, t)$ and a supply curve
$W_j(x, t)$, it is natural to ask: {\em What is the cost of acquiring a
given supply $s_j$ of type $j$?}.  Using the monotonicity of $W_j$,
the lowest bid necessary to obtain $s_j$ supply is
$x_j = \text{min}\{x \in \R\; |\; W_j(x) \ge s_j\}$.  After applying
randomized bidding, since $W_j$ is then strictly monotone, this is
simply $x_j = W_j^{-1}(s_j, t)$, where the inverse is w.r.t. $x$.  The
cost of acquiring $s_j$ units of type $j$ instantaneously at time $t$,
which we will denote by $\Lambda_j(s_j, t)$, is therefore
\begin{equation}
  \label{eqn:production_cost}
  \Lambda_j(s_j, t) \defeq f_j(W_j^{-1}(s_j, t), t).
\end{equation}

For second price auctions, this function turns out to be convex.  We
suppress the $t$ argument and the index $j$ in the following.

\begin{prop}[Convex Acquisition Costs]
  \label{prop:convex_acquisition_costs}
  In a second price auction, the acquisition function
  $\Lambda(s) = f \circ W^{-1}(s)$ is convex.  Moreover,
  $\Lambda(0) = 0,$ and $\Lambda'(s) = \ind[s \ge W(0)]W^{-1}(s)$.
\end{prop}

\noindent{\bf Proof:}

  Consider the integral representation of the cost function in
  Equation \eqref{eqn:sp_cost} and make the substitution
  $y = W(u) \implies \d y = W'(u) \d u$ to obtain
  \begin{align*}
    \Lambda(x)
    &= \ind[W^{-1}(x) \ge 0]\int_0^{W^{-1}(x)} u W'(u)\d u\\
    &= \ind[x \ge W(0)]\int_{W(0)}^xW^{-1}(y)\d y.
  \end{align*}

  It is seen here $\Lambda(x) = 0$ for $x \le W(0)$, in particular,
  $\Lambda(0) = 0$.  Moreover, this function is differentiable on
  $x > W(0)$ and $\Lambda'(x) = W^{-1}(x)$.  On $x < W(0)$, we have
  $\Lambda'(x) = 0$.  Since $W^{-1}(W(0)) = 0$, $\Lambda$ is
  continuously differentiable on $\R$.  Since $W$ is monotone, so is
  $W^{-1}$, and it is well known that functions with monotone
  derivatives are convex.

  An analogous proof can be given for non-smooth (even discontinuous)
  supply curves to show that $\Lambda_j$ is convex and differentiable by
  using the generalized inverse
  $$W^{-1}(x) = \text{min}\{y \in \R\ |\ W(y) \ge x\},$$
  and the substitution rule for the Lebesgue-Stieljes
  integral (see e.g. \cite{falkner2012substitution}).

\begin{remark}[General Supply Curves]
  The work of \cite{marbach_bidding_2020} studies the second price
  auction in the case where $W$ may be any right-continuous and
  non-decreasing supply curve.  Proposition
  \ref{prop:convex_acquisition_costs} does not rely on the smoothness
  of $W$ and in fact applies to this more general case.
\end{remark}

\section{Time Constrained Impression Contracts}
\label{sec:time_constrained_impression_contracts}
We consider a DSP tasked with managing contracts (or ``campaigns'') of
the form $\mC = (T, C, S)$, where $T \in \R_{++}$ is a time deadline,
$C \in \N$ is the number of items that must be won in auction by the
deadline, and $S$ is a set of targeting criteria specifying the
characteristics of impressions that can be used to satisfy the terms
of the contract.

Suppose that we have a finite set $\Omega$ of possible impression
characteristics (i.e.\ sex, age, location, publisher, etc.), where we
note that our DSP is essentially free to construct this set.  For
instance, we may have $\Omega = \{0, 1\}^L$ where each dimension
indicates the presence or absence of a particular characteristic.  We
will allow for any set $S \subseteq \Omega$ which satisfies some
natural consistency rules (e.g.,
$S = \{\mathsf{sex} = \mathsf{female}, \mathsf{sex} \ne
\mathsf{female}\}$ is inconsistent) to be associated with a contract.
Then, any impression with characteristics $I \in \Omega$ won in an
RTB auction is allowed to count towards satisfying the contract if
$I \in S$.  That is, if $I$ \textit{matches} the type
specification given by $S$.

\subsection{Target Criteria Decomposition}
\label{sec:target_partitioning}
In this section, we discuss a target set partitioning important for
the formulation of our main problem as in \cite{marbach_bidding_2020}.

Suppose that we have $N$ contracts $\{\mC_i\}_{i = 1}^N$, where the
targeting sets $\mS \defeq \{S_i\}_{i = 1}^N$ may be overlapping.  

It is clear that there exists some minial $M$ and disjoint sets $R_j$
such that:

\begin{equation}
\label{eq:target}
\bigcup_{j=1}^M R_j = \bigcup_{i=1}^N S_i.
\end{equation}

Moreover for each $i\in [N]$ there exists a \textit{unique} set
$\A_i \subseteq [M]$ such that:

\begin{equation}
\bigcup_{j \in \A_i} R_j = S_i.
\end{equation}

This in turn also induces a set $\B_j \subseteq [N]$ such that
$$i \in \B_j \iff R_j \subseteq S_i.$$  
And, moreover,
$$j \in \A_i \iff i \in \B_j.$$

\begin{figure}[h!]
  \centering
  \caption{Set Partitioning Example}
  \label{fig:set_partitions}
  \includegraphics[width=0.75\textwidth]{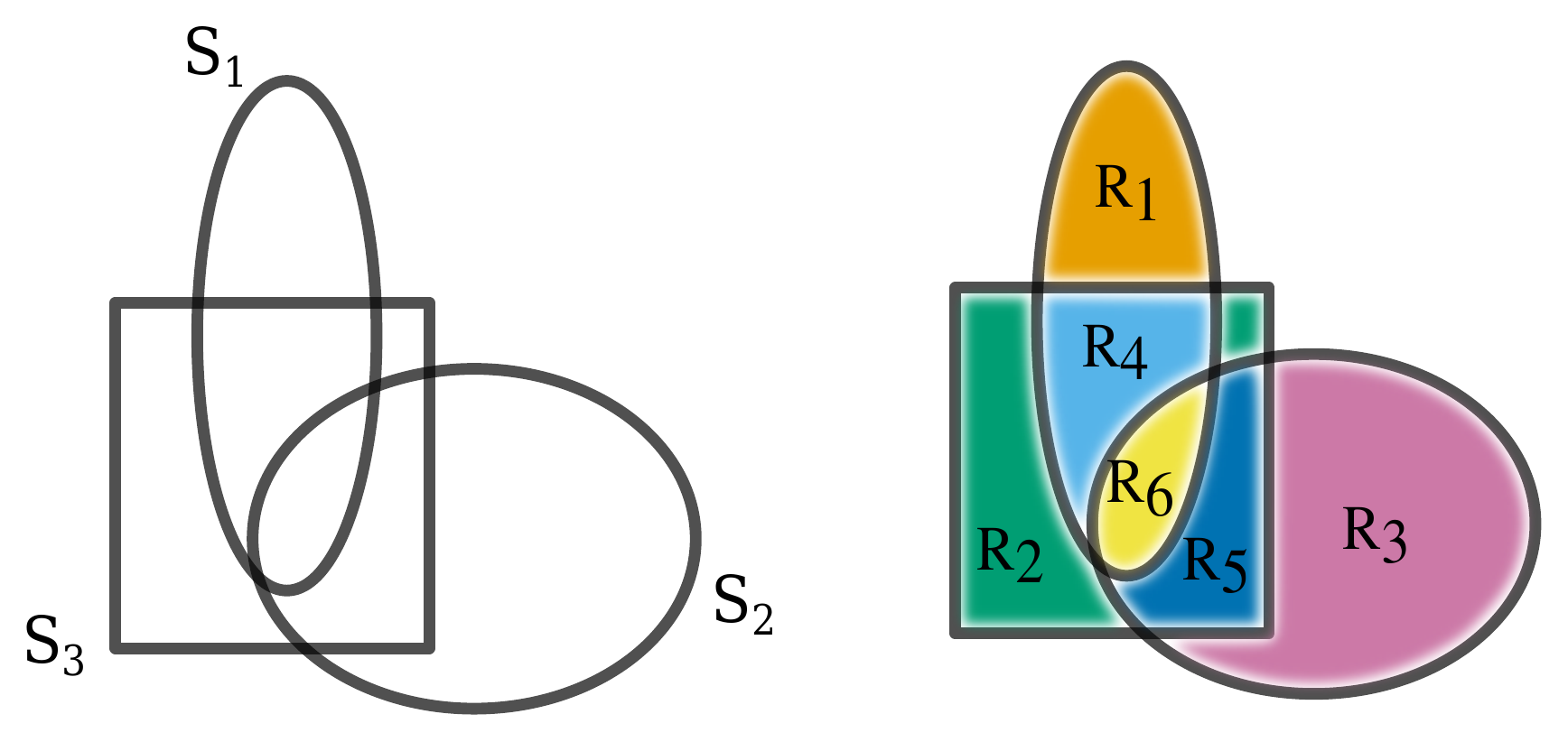}

  \footnotesize{An example of set partitioning, best viewed in colour.  In
    this case, $\mS = \{S_1, S_2, S_3\}$, $M = 6$ and
    $\mR = \{R_m\}_{m = 1}^M$ contains subsets such that $\mR$ is a
    partition of $\bigcup \mS$.  Moreover, for any $S_i \in \mS$ we
    have some $\mA_i \subseteq [M]$ such that
    $\bigcup_{j \in \mA_i} R_j = S_i.$ For example,
    $S_2 = R_3 \cup R_5 \cup R_6$.  That is, $\mA_2 = \{3, 5, 6\}$.
    Likewise, we have sets $\mB_j$ such that
    $j \in \mA_i \iff i \in \mB_j$.  For example, $\mB_1 = \{1\}$ and
    $\mB_6 = \{1, 2, 3\}$.}
\end{figure}

The interpretation is that $R_j$ represents a targeting criterion
while $B_j$ is the set of campaigns that require impressions
satisfying criteria $R_j$.  With this decomposition the supply curve
$W_j(t,x)$ will denote the supply curve for the impressions that match
$R_j$.  An example of such a partition is provided by Figure
\ref{fig:set_partitions}.

\section{Optimal Management of Impression Contracts: Preliminaries}
\label{sec:optimal_control}
In this section we will formulate optimal control problems for
fulfilling impression contracts.  We begin with the simplest case
where there is a single type of item (we don't distinguish between bid
requests, or, $M = 1$) and a single contract (i.e., $N = 1$)
stipulating that we must obtain $C$ impressions by time $T$.  We will
start with a time-homogeneous problem wherein $W(x, t) = W(x)$ for
every $t$, and similarly for the cost.  The simplicity ensures that
this problem has what is essentially a closed form solution.  We use
this example to illustrate methods for revising the bid over time as
more information becomes available via a receding horizon.

\begin{remark}[Notation]
  For the optimization problems presented in this paper, we follow a
  convention for constraints where indices (e.g.\ $i, j, t$) that do
  not appear explicitly in summation or integration indicate that
  there is one constraint for every combination of valid indices.  For
  example,
  $$\gamma_{ij}(t) \ge 0; \forall i \in [N], \forall j \in \A_i, \forall
  t \in [0, T],$$ will be written simply as $\gamma_{ij}(t) \ge 0$.
  Some attention must also be given to the combinations of indices
  which are valid, e.g., we do not refer to any $\gamma_{ij}$ for
  which $j \not\in \A_i$.
\end{remark}

\subsection{A Single Item Type ($M = 1$)}
\label{sec:single_contract_single_supply}
We begin with the case where we are obliged simply to fulfill a single
contract $(T, C, S)$.  Firstly, suppose that the structure of $S$ is
simple enough that we are satisfied with the estimate of a single
supply curve $W(x, t)$, i.e., all items satisfying $S$ are estimated
as having the same $W$.  Furthermore, suppose for now that the supply
curve does not depend on $t$, i.e., $W(x, t) = \lambda W(x)$.  We will
see later that this assumption is not restrictive.  The function here
$W(x)$ is now a bonafide cumulative distribution function, and
$\lambda > 0$ is the average rate of supply.  Our problem is then
\begin{equation}
  \label{eqn:simple_case}
  \begin{aligned}
    \underset{x}{\text{minimize}} \quad & \int_0^Tf(x(t)) \d t\\
    \textrm{subject to}
    \quad & \lambda \int_0^T W(x(t))\d t \ge C.
  \end{aligned}
\end{equation}

Making the substitution $s(t) = W(x(t))$ we can rewrite this as a
convex problem:
\begin{equation}
  \begin{aligned}
    \underset{s}{\text{minimize}} \quad & \int_0^T\Lambda(s(t)) \d t\\
    \textrm{subject to}
    \quad & \lambda \int_0^T s(t)\d t \ge C,
  \end{aligned}
\end{equation}

where we recall that $\Lambda = f \circ W^{-1}$.  This is a classical
calculus of variations problem with integral constraints (see
\cite[Theorem~14.12]{clarke2013functional}).  We have the Lagrangian
with $\mu \geq 0$:

$$\mathcal{L}(s, \mu) = \Lambda(s) - \mu\bigl[\lambda s - C\bigr].$$

Any $\mathcal{C}_1$ solution $s(t)$ necessarily satisfies the
Euler-Lagrange equation

$$\Lambda'(s(t)) = \mu\lambda\; \forall\ t \in [0, T].$$

Since $\Lambda' = W^{-1}$, it is necessary that
$x(t) = \mu \lambda.$ That is, $x(t)$ must be a constant.

Substituting this into the cost and impression constraints given by
(\ref{eqn:simple_case}), we see that since $f(x)$ and $W(x)$ are
monotone increasing functions, the optimal $x$ is the smallest
feasible bid:
\begin{equation}
  \label{eqn:x_star_simplest}
  x^\star(C, T) = \left\{
    \begin{array}{lr}
      W^{-1}\big(\frac{C}{\lambda T}\big) & \frac{C}{\lambda T} < 1\\
      \underset{x \rightarrow 1}{\text{lim}}\  W^{-1}(x) & \text{otherwise}
    \end{array}\right..
\end{equation}

We define $x^\star$ through a ``best effort'' limit if the problem is
not feasible in order to define a complete bidding strategy for the
DSP.  Note also that the inverse of $W$ is guaranteed to exist by the
\textit{strict} monotonicity of $W$, see Definition
\ref{def:supply_curve}.

\subsubsection{Receding Horizon Control.}
\label{sec:receding_horizon}
The bid path \eqref{eqn:x_star_simplest} does not take into account
any of the information gained during the course of bidding.  In this
case, it is natural to convert our solution into a receding horizon
(RH) (see \cite{cannon2016model}) algorithm where if after time $t$ has
elapsed, we have accumulated $c(t)$ supply, we can modify the
constraints of the problem from $C$ to $C - c(t)$ and the constant $T$
to $T - t$, resulting in the RH control algorithm
\begin{equation}
  \label{eqn:receding_horizon_example}
  x_{\text{rh}}(t) \defeq x^\star(C - c(t), T - t) = \left\{
    \begin{array}{lr}
      W^{-1}\big(\frac{1}{\lambda}\frac{C - c(t)}{T - t}\big) & \frac{1}{\lambda}\frac{C - c(t)}{T - t} < 1\\
      \underset{x \rightarrow 1}{\text{lim}}\ W^{-1}(x) & \text{otherwise}
    \end{array}\right..
\end{equation}

The RH framework accounts for unexpected supply shortages or surpluses
and also enables us to naturally incorporate a case wherein new
contracts arrive before the set of current contracts have been
fulfilled.

  \begin{example}
    We consider an illustrative example where the DSP forecasts supply
    with the parametric form $W(x) = 1 - e^{-\gamma x}$, and constant
    supply $\lambda_0$.  Clearly,
    $W^{-1}(s) = -\frac{1}{\gamma}\ln (1 - s)$ for $s \in [0, 1)$, from which
    the optimal bids, including receding horizon are immediately
    derived
  \begin{equation}
    \label{eqn:RH_simple_example}
    x_{\text{rh}}(t) = \left\{
      \begin{array}{lr}
        -\frac{1}{\gamma} \ln \bigl[1 - \frac{1}{\lambda_0}\frac{C - c(t)}{T - t} \bigr]& \frac{1}{\lambda_0}\frac{C - c(t)}{T - t} < 1\\
        \infty & \text{otherwise}
      \end{array}\right..
  \end{equation}

  Suppose now that the realized supply over the period $[0, T]$ obeyed
  the law $\lambda(t) W(x)$, i.e., the DSP's estimate of supply is in
  error by $\lambda(t) - \lambda_0$.  Figure \ref{fig:control_example}
  illustrates the behaviour of the static and receding horizon
  algorithms for the case of undersupply:
  $\frac{1}{T}\int_0^T \lambda(t)\d t < \lambda_0$, and oversupply
  $\frac{1}{T}\int_0^T \lambda(t)\d t > \lambda_0$.

  For the receding horizon case, the supply \textit{actually} attained
  can be described by the differential equation
  \begin{equation}
    \label{eqn:rh_ode_example}
    \dot{c}_{\text{rh}}(t) = \lambda(t)W(x^{\star}(C - c_{\text{rh}}(t), T - t));\ c_{\text{rh}}(0) = 0,
  \end{equation}

  and the analogous equations for the static case $c(t)$.  Since the
  optimal bid $x^{\star}$ involves the inverse of the win probability
  $W^{-1}$, substituting it into Equation \eqref{eqn:rh_ode_example}
  results in a separable ordinary differential equation
  \begin{align*}
    \dot{c}_{\text{rh}}(t) &= \frac{\lambda(t)}{\lambda_0}\frac{C - c_{\text{rh}}(t)}{T - t};\; c_{\text{rh}}(0) = 0\\
    \implies c_{\text{rh}}(t) &= C\Bigl[1 - \text{exp}\bigl(-\frac{1}{\lambda_0}\int_0^t \frac{\lambda(s) \d s}{T - s}\bigr) \Bigr],
  \end{align*}

  which reduces simply to the straight line
  $c_{\text{rh}}(t) = \frac{Ct}{T}$ if the estimate is accurate and
  $\lambda(t) = \lambda_0$.  The intuition that good acquisition paths
  are simply straight lines when prices are time-independent is
  reinforced by examining the curves and their relative costs in
  Figure \ref{fig:control_example}.

  This solution corresponds to estimates of an \textit{average}
  behaviour, further simulation results including discrete event
  simulations with real market data are developed in section
  \ref{sec:simulations}.

  \begin{figure}
    \centering
    \caption{Receding Horizon Acquisition Paths}
    \label{fig:control_example}

    \begin{subfigure}[b]{0.4\textwidth}
      \caption{Undersupply}
      \includegraphics[width=\textwidth]{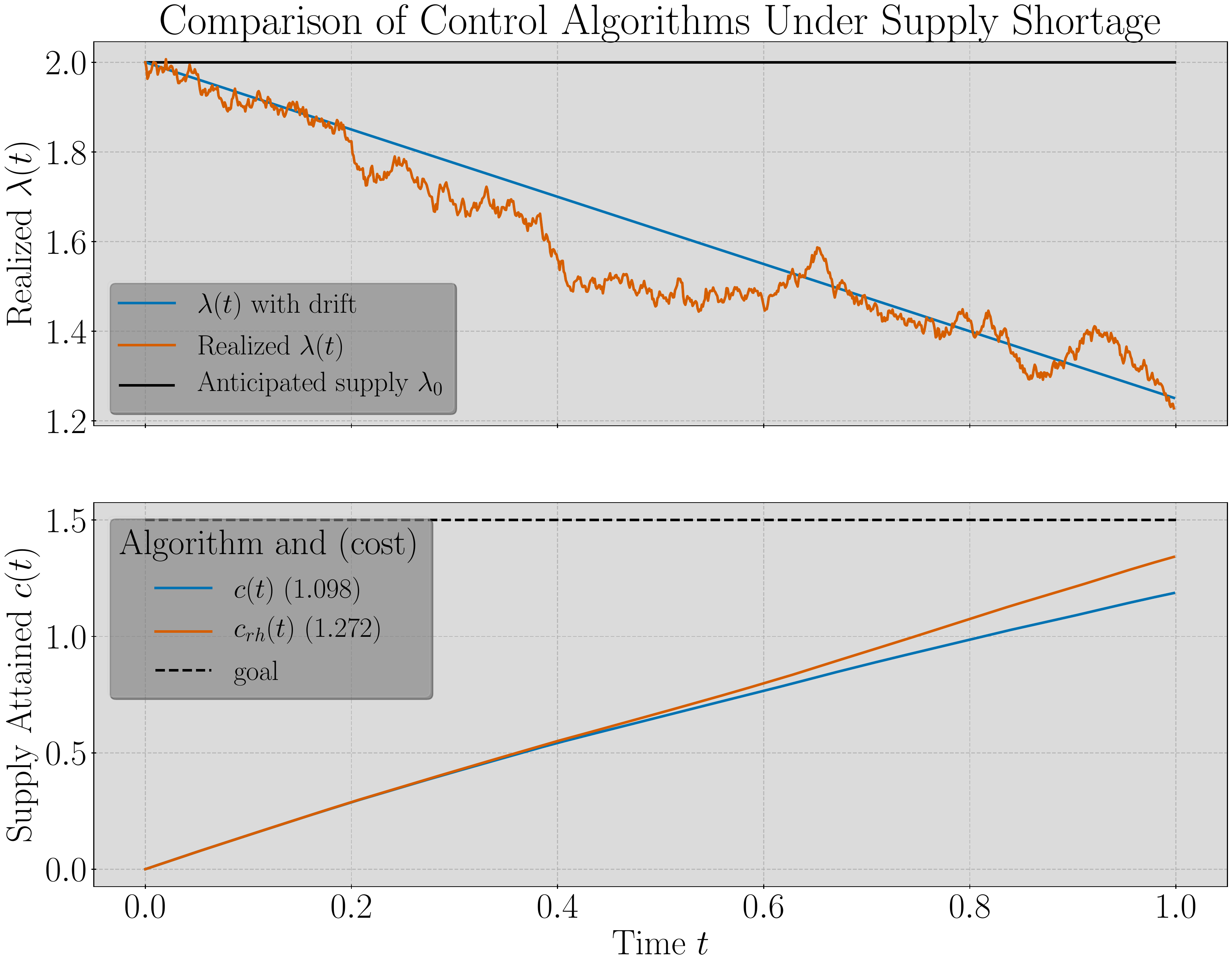}
    \end{subfigure}
    \begin{subfigure}[b]{0.4\textwidth}
      \caption{Oversupply}
      \includegraphics[width=\textwidth]{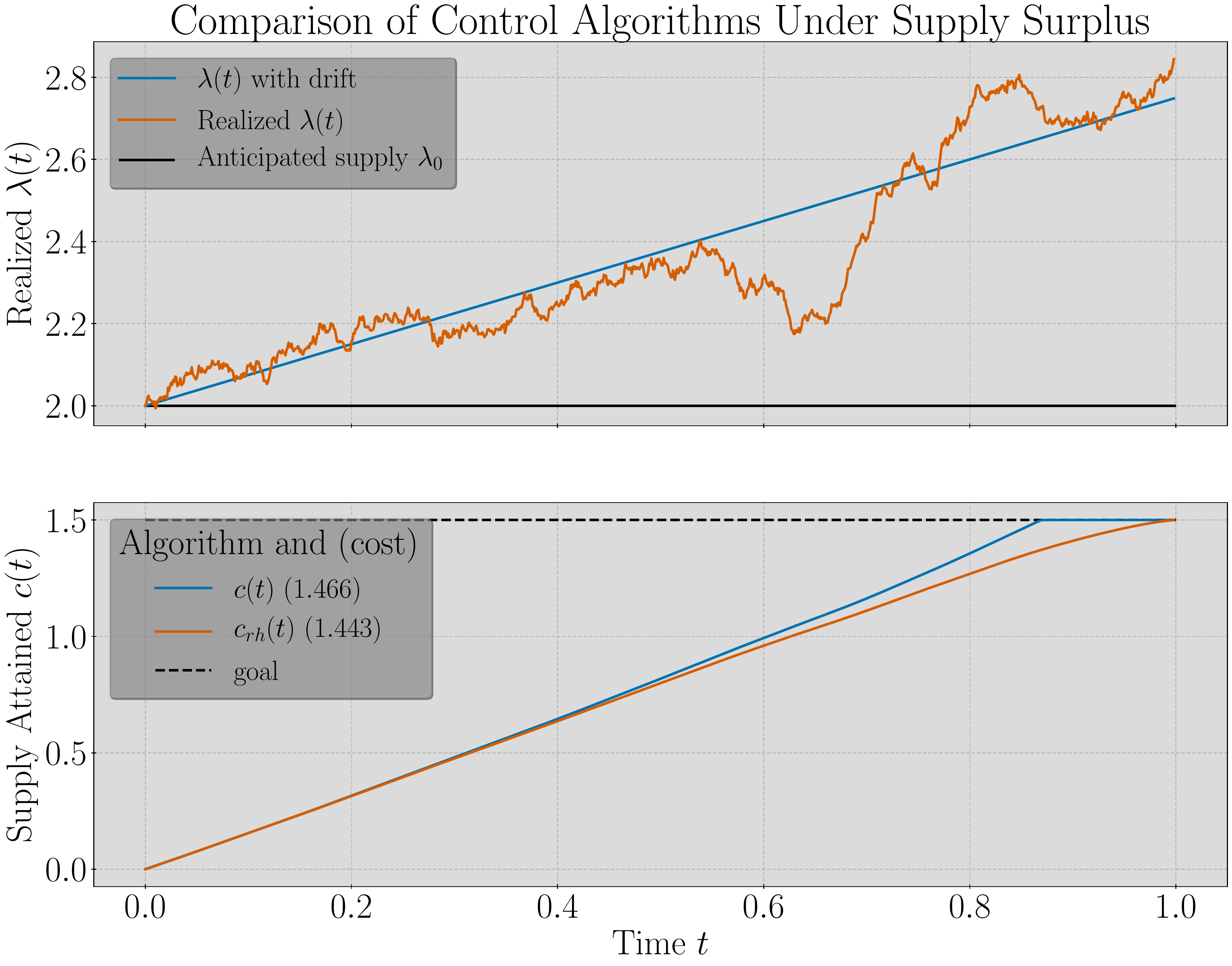}
    \end{subfigure}

    \footnotesize{Simulated acquisition paths $c(t)$ for the case
      $M = N = 1$ comparing the behaviour of different algorithms in
      the presence of supply shortages or surpluses in comparison to
      expectation $\lambda_0$.  Best viewed in colour.  Qualitatively,
      when there is oversupply, the receding horizon smooths the
      acquisition rate to reduce costs, and when there is
      undersupply, it increases the bid in reaction to the shortage.}
  \end{figure}
\end{example}

\section{Optimal Management of Impression Contracts}
\label{sec:general_case}
Following the decomposition of Section \ref{sec:target_partitioning},
we have a collection of $N$ contracts indexed by $i$ with differing
deadlines $0 < T_1 \le \cdots \le T_N \defeq T$ and another set of $M$
item types indexed by $j$.  This induces a problem where we need to
calculate an \textit{array} of bids $x(t) \in \R_+^{N \times M}$, as
well as an \textit{array} of allocations
$\gamma(t) \in [0, 1]^{N \times M}$.  The interpretation is that if an
item of type $j$ arrives at time $t$ the quantity
$\sum_{i \in \B_j}\gamma_{ij}(t)$ indicates the probability of bidding
on the item, and $\gamma_{ij}(t)$ is the probability of allocating
that item (if won) to fulfill contract $i$.  The bid which is submit
is given by $x_{ij}(t)$.

We naturally have the constraints $\gamma_{ij}(t) = 0$ if
$i \notin \B_j$, or equivalently $j \notin \A_i$.  Indeed, we may
think of $\gamma$ as weights on the edges of a bipartite graph with
nodes $[N] \times [M]$ and an edge $(i, j)$ if $i \in \B_j$.

The contract deadlines are an important detail of the problem, and
induce a set of times $T^1, \ldots T^M$ for the item types where
$T^j \defeq \underset{i \in \B_j}{\text{max}}\; T_i$ is the last instant
that an item of type $j$ is useful.  Moreover, we will see that the
sets $\Tt \defeq \{i \in [N]\; |\; t < T_i\}$ of contracts active up
to (but not including) time $t$ and the set
$\Ttj \defeq \{j \in [M]\; |\; t < T^j\}$ of items useful up to (but
not including) time $t$ will arise naturally.

Finally, since $\gamma_{ij}(t)$ is the allocation proportion from $j$
to $i$, we must have that $\gamma_{ij}(t) = 0$ for any $t \ge T_i$ or
$t \ge T^j$ and that $\sum_{i \in \B_j} \gamma_{ij}(t) \le 1$.  In
fact, we can see that, necessarily,
$\sum_{i \in \B_j \cap \Tt}\gamma_{ij}(t) \le \ind[t < T^j]$.

We formulate the joint problem for fulfilling the contracts as an
optimal control problem as follows:
\begin{equation}
  \label{eqn:ct_problem}
  \begin{aligned}
    \underset{x, \gamma}{\text{minimize}} \quad & \sum_{i = 1}^N\int_{0}^{T_i} \Bigl[\sum_{j \in \A_i \cap \Ttj}\gamma_{ij}(t)f_j(x_{ij}(t), t)\Bigr]\d t\\
    \textrm{subject to} \quad & \dot{c}_i(t) = \ind[t < T_i]\sum_{j \in \A_i \cap \Ttj}\gamma_{ij}(t)W_j(x_{ij}(t), t)\\
    \quad & \sum_{i \in \B_j \cap \Tt} \gamma_{ij}(t) \le \ind[t < T^j]\\
    \quad & c_i(0) = 0, c_i(T) \ge C_i, \gamma_{ij}(t) \ge 0,
  \end{aligned}
  \tag{$P$}
\end{equation}

where the state $c_i(t)$ indicates the expected supply obtained by
time $t$ for contract $i$.

This is a direct generalization of \cite{marbach_bidding_2020} to the
case where supply curves are time-dependent, and crucially, where
there may be differing contract deadlines.

\begin{remark}[Single Contract, Multiple Item Types $(N = 1, M > 1)$]
  The special case of \eqref{eqn:ct_problem} when $N = 1$ may be of
  interest since the $N > 1$ case could be approached by solving $N$
  instances of this special case.  However, this would have the
  obvious drawback of putting the DSP in competition with itself.
\end{remark}

\subsection{The Convex Reformulation}
\label{sec:convex_reformulation}
Since the cost of acquisition function $\Lambda_j$ is convex, it
suggests that Problem \eqref{eqn:ct_problem} can be reformulated into
a convex problem.  In order to carry out this transformation, we first
show that the bids $x_{ij}(t)$ can be chosen independently of $i$,
that is, $x_{ij}(t) = x_j(t)$.

\begin{prop}[Uniform Bid Principle (UBP)]
  \label{prop:ubp}
  Any solution $(x, \gamma)$ of Problem \eqref{eqn:ct_problem} can be
  transformed into another solution $(\tilde{x}, \tilde{\gamma})$ such
  that $\tilde{x}_{uj}(t) = \tilde{x}_{vj}(t)$ for every
  $u, v \in [N]$, and moreover, such that
  $\forall t < T^j\; \sum_{i \in \B_j \cap \Tt}\tilde{\gamma}_{ij}(t) \in \{0, 1\}$.
\end{prop}

\noindent{\bf Proof:} 

  Suppose $(x, \gamma)$ is a solution of Problem
  \eqref{eqn:ct_problem}, and with total cost $J$. Let
   $(\tilde{x}, \tilde{\gamma})$ be another solution with total cost
  $\tilde{J}$ and where
  \begin{align*}
    \tilde{x}_j(t)
    &\defeq W_j^{-1}\Bigl(\sum_{i \in \B_j\cap\Tt}\gamma_{ij}(t)x_{ij}(t), t\Bigr),\\
    \tilde{\gamma}_{ij}(t)
    &\defeq \frac{\gamma_{ij}(t) W_j(x_{ij}(t), t)}{\sum_{u \in \B_j\cap \Tt} \gamma_{uj}(t) W_j(x_{uj}(t), t)},
  \end{align*}
  where $0 / 0 \defeq 0$ in the definition of $\tilde{\gamma}$.

  It is clear that $\tilde{\gamma}_{ij}(t)$ is feasible since
  $\tilde{\gamma}_{ij}(t) \ge 0$ and
  $\sum_{i \in \B_j \cap \Tt} \tilde{\gamma}_{ij}(t) \le \ind[t <
  T^j]$ by definition.  Indeed,
  $\forall t < T^j\; \sum_{i \in \B_j \cap \Tt}\tilde{\gamma}_{ij}(t)
  \in \{0, 1\}$.

  The cost of $(\tilde{x}, \tilde{\gamma})$, instantaneously at time
  $t$, then satisfies $\tilde{J} = J$ since $J$ is the minimal cost and
  \begin{align*}
    \tilde{J}
    &\defeq \sum_{i \in \Tt} \sum_{j \in \A_i} \tilde{\gamma}_{ij}(t) f_j(\tilde{x}_j)\\
    &\overset{(a)}{=} \sum_{i \in \Tt} \sum_{j \in \A_i} \tilde{\gamma}_{ij}(t) \Lambda_j\Bigl(\sum_{u \in \B_j \cap \Tt}\gamma_{uj}(t)W_j(x_{uj}(t), t), t\Bigr)\\
    &\overset{(b)}{\le} \sum_{i \in \Tt} \sum_{j \in \A_i} \tilde{\gamma}_{ij}(t) \sum_{u \in \B_j \cap \Tt}\gamma_{uj}(t) \Lambda_j(W_j(x_{uj}(t), t), t)\\
    &\overset{(c)}{=} \sum_{j \in \Ttj}\sum_{u \in \B_j \cap \Tt} \gamma_{uj}(t) f_j(x_{uj}(t), t) \sum_{i \in \B_j \cap \Tt} \tilde{\gamma}_{ij}(t) = J\\
  \end{align*}

  where $(a)$ is just the definition of $\Lambda_j$ (c.f. Proposition
  \ref{prop:convex_acquisition_costs}), $(b)$ follows by the convexity
  of $\Lambda_j$ and that $\Lambda_j(0) = 0$ (since $\gamma_{ij}$ need
  not necessarily sum to $1$), and $(c)$ since
  $\Lambda_j = f_j \circ W_j^{-1}$ and then by swapping the order of
  summation using $i \in \B_j \iff j \in \A_i$.
\vspace{0.4cm}

With this proposition in hand, there is no reason to consider
solutions where the bids depend on $i$.  This fact enables us to make
significant simplifications to Problem \eqref{eqn:ct_problem}.  Rather
than optimizing over the bid and allocation pair $(x, \gamma)$, we can
instead optimize over a supply and unnormalized allocation $(s, r)$
where $s_j(t) = W_j(x_j(t), t)$ and
$r_{ij}(t) = \gamma_{ij}(t) s_j(t)$.  We summarize this idea in the
following proposition, with a detailed description of the
transformation provided in the Appendix.

\begin{prop}[Convex Formulation]
  Problem \eqref{eqn:ct_problem} can be equivalently reformulated as
  the following convex optimization problem
  \begin{equation}
    \label{eqn:ct_problem_cvx}
    \begin{aligned}
      \underset{s, r}{\mathrm{minimize}} \quad & \sum_{j = 1}^M \int_0^{T^j}\Lambda_j(s_j(t), t) \d t\\
      \mathrm{subject\; to} \quad & \dot{c}_i(t) = \ind[t < T_i]\sum_{j \in \A_i}r_{ij}(t)\\
      \quad & \sum_{i \in \B_j \cap \Tt} r_{ij}(t) = s_j(t) \ind[t < T^j]\\
      \quad & c_i(0) = 0, c_i(T) \ge C_i, r_{ij}(t) \ge 0.
    \end{aligned}
    \tag{$P_{\text{cvx}}$}
  \end{equation}

  A solution to the original problem is obtained via
  $x_j(t) = W_j^{-1}(s_j(t), t)$ and
  $\gamma_{ij}(t) = r_{ij}(t) / s_j(t)$.
\end{prop}

\noindent{\bf Proof:}

  Recall the original problem \eqref{eqn:ct_problem}, and apply
  Proposition \ref{prop:ubp} to eliminate the dependence of the bid on
  $i$:
  \begin{equation}
    \begin{aligned}
      \underset{x, \gamma}{\text{minimize}} \quad & \sum_{i = 1}^N\int_{0}^{T_i} \Bigl[\sum_{j \in \A_i \cap \Ttj}\gamma_{ij}(t)f_j(x_j(t), t)\Bigr]\d t\\
      \textrm{subject to} \quad & \dot{c}_i(t) = \ind[t < T_i]\sum_{j \in \A_i \cap \Ttj}\gamma_{ij}(t)W_j(x_j(t), t)\\
      \quad & \sum_{i \in \B_j \cap \Tt} \gamma_{ij}(t) \le \ind[t < T^j]\\
      \quad & c_i(0) = 0, c_i(T) \ge C_i, \gamma_{ij}(t) \ge 0.
    \end{aligned}
    \tag{$P$}
  \end{equation}

  Due to the bid's independence of $i$, we can rearrange the objective
  by swapping the order of summation:
  \begin{equation}
    \sum_{i = 1}^N\int_{0}^{T_i} \Bigl[\sum_{j \in \A_i \cap \Ttj}\gamma_{ij}(t)f_j(x_j(t), t)\Bigr]\d t
    = \sum_{j = 1}^M\int_0^{T^j}\Bigl[f_j(x_j(t), t) \sum_{i \in \B_j \cap \Tt}\gamma_{ij}(t)\Bigr]\d t,
  \end{equation}

  which, after making the substitution $s_j(t) = W_j(x_j(t), t)$, results in
  \begin{equation}
    \begin{aligned}
      \underset{s, \gamma}{\text{minimize}} \quad & \sum_{j = 1}^M\int_{0}^{T^j} \Bigl[\Lambda_j(s_j(t), t) \sum_{i \in \B_j \cap \Tt}\gamma_{ij}(t)\Bigr]\d t\\
      \textrm{subject to} \quad & \dot{c}_i(t) = \ind[t < T_i]\sum_{j \in \A_i \cap \Ttj}\gamma_{ij}(t)s_j(t)\\
      \quad & \sum_{i \in \B_j \cap \Tt} \gamma_{ij}(t) \le \ind[t < T^j]\\
      \quad & c_i(0) = 0, c_i(T) \ge C_i, \gamma_{ij}(t) \ge 0.
    \end{aligned}
  \end{equation}

  Now, make the substitution $r_{ij}(t) \defeq \gamma_{ij}(t)s_j(t).$
  Notice that Proposition \ref{prop:ubp} also ensures that we have a
  solution where
  $\sum_{i \in \B_j \cap \Tt} \gamma_{ij}(t) \in \{0, 1\}$, and if
  this summation is $0$, then necessarily $s_j(t) = 0$ which in turn
  implies that $\Lambda_j(s_j(t), t) = 0$.  Therefore we can write
  \begin{equation}
    \begin{aligned}
      \underset{s, \gamma}{\text{minimize}} \quad & \sum_{j = 1}^M\int_{0}^{T^j} \Lambda_j(s_j(t), t)\d t\\
      \textrm{subject to} \quad & \dot{c}_i(t) = \ind[t < T_i]\sum_{j \in \A_i \cap \Ttj} r_{ij}(t)\\
      \quad & \sum_{i \in \B_j \cap \Tt} r_{ij}(t) = s_j(t) \ind[t < T^j]\\
      \quad & c_i(0) = 0, c_i(T) \ge C_i, r_{ij}(t) \ge 0.
    \end{aligned}
    \tag{$P_{\text{cvx}}$}
  \end{equation}

  Problem \eqref{eqn:ct_problem_cvx} has linear constraints and a
  convex objective, and is therefore itself a convex optimization
  problem.
\vspace{0.3cm}

\begin{remark}
  In this formulation, we see a close connection to the
  Production-Transportation problem \cite{leblanc1974transportation,
    sharp1970decomposition}, where $\Lambda_j$ are the production
  costs, and the transportation costs belong to the set $\{0, \infty\}$
  encoding the set $\A_i, \B_j$.  The proof of Proposition
  \eqref{prop:convex_acquisition_costs} is essentially establishing
  that the \textit{marginal} production costs, $\Lambda_j'$, are
  monotone: a key aspect of the analysis of
  \cite{leblanc1974transportation, sharp1970decomposition}.
\end{remark}

\subsection{Necessary Conditions}
\label{sec:necessary_conditions}
We characterize the necessary properties of $(s(t), r(t))$ (and
$(x(t), \gamma(t))$ by extension) via the Pontryagin Maximum Principle
(\cite{clarke2013functional}).  Define the Hamiltonian
\begin{equation}
  \label{eqn:sc_ct_problem_hamiltonian}
  \H(t, p, c, s, r) = \sum_{i \in \Tt}p_i \sum_{j \in \A_i}r_{ij} - \sum_{j \in \Ttj}\Lambda_j(s_j, t).
\end{equation}

The question of constraint qualifications and the existence of
solutions is of technical importance, but we will defer these issues
to Section \ref{sec:existence_and_optimality}, assuming for now
that a solution sufficiently regular to allow the application of the
maximum principle does, in fact, exist.

The maximum principle ensures that there exists some absolutely
continuous function $p: [0, T] \rightarrow \R^N$ that satisfies the
\textit{adjoint equation}
\begin{equation}
  \dot{p}(t)^\T = -\D_c\H(t, p, c, s, r)
\end{equation}

But, since $\H$ does not depend on $c$ explicitly, $\dot{p}(t) = 0$,
and the adjoint is a constant $p \in \R^N$.  Denote
$E_i = [C_i, \infty)$ and
\begin{equation}
  \label{eqn:normal_cone}
  N_{E_i}(x) = \left\{
    \begin{array}{cc}
      \R_- & x = C_i\\
      \{0\} & x > C_i\\
    \end{array}
    \right.,
\end{equation}

the normal cone.  The maximum principle requires
$-p_i(T) \in N_{E_i}(c_i(T))$ and therefore that for any optimal state
$c_i(t)$, we must have $p_i = 0$ for $c_i(T) > C_i$ (i.e., in the case
of over fulfillment) and $p_i \ge 0$ for $c_i(T) = C_i$.  Ultimately,
this implies that $p_i \ge 0$.

\begin{remark}[Pathological Cases]
  The unusual case of an optimal solution satisfying $c_i(T) > C_i$ is
  in fact possible.  This may arise from the randomized bidding and
  that we may have $W_j(0) > 0$ even though $f_j(0) = 0$.  If the
  supply requirements $C_i$ are extremely small, then our model allows
  the attainment of this supply at $0$ cost.  This is an artifact of
  the technical assumptions necessary to rigorously establish our
  results, but is not of practical relevance: for bids well within the
  interior of $\R_+$, the approximation error in the cost function is
  negligible.
\end{remark}

Finally, a solution must satisfy the maximum condition
\begin{equation}
  \H(t, p, c(t), s(t), r(t)) = \underset{(s, r) \in U(t)}{\text{sup}}\; \H(t, p, c(t), s, r),
\end{equation}

where $U(t)$ encodes the constraints.  Since we are already asserting
the attainment of the above suprema, we can formulate the problem of
extremizing the Hamiltonian at time $t$:
\begin{equation}
  \label{eqn:sc_hamiltonian_problem}
  \begin{aligned}
    \underset{s, r}{\text{maximize}} \quad & \sum_{i \in \Tt}\sum_{j \in \A_i} p_ir_{ij} -  \sum_{j \in \Ttj}\Lambda_j(s_j, t)\\
    \textrm{subject to}
    \quad & \sum_{i \in \B_j \cap \Tt}r_{ij} = s_j(t), r_{ij} \ge 0.\\
  \end{aligned}
\end{equation}

We will see that there is a tight relationship between the bids
$x_j(t)$ and the adjoint vector $p$, such that the entire continuous
time path $x(t)$ will be fully determined by the finite vector $p$ --
for this reason, we refer to $p$ as the vector of
\textit{pseudo-bids}.  Moreover, this pseudo-bid vector determines
some key aspects of the \textit{support} (i.e., indices of non-zero
entries) of $r_{ij}(t).$ Introducing notation for the maximum
pseudo-bid over the set $\B_j \cap \Tt$
\begin{equation}
  p_j^\star(t) \defeq \underset{i \in B_j \cap \Tt}{\text{max}} p_i,
\end{equation}

we have the following proposition:

\begin{prop}[Optimal Allocation]
  \label{prop:optimal_allocation}
  Any (regular) solution $(r, s)$ of \eqref{eqn:ct_problem_cvx} and
  the corresponding acquisition path $c_i(t)$ and vector of
  pseudo-bids $p$ must satisfy $c_i(T_i) \ge C_i$ and $p_i \ge 0$ for
  every $i$.  Moreover, $(s, r)$ maximizes the Hamiltonian at time $t$
  if and only if
  \begin{subequations}
    \begin{align}
      i\in \B_j \cap \Tt, p_i < p_j^\star(t) &\implies r_{ij}(t) = 0. \label{eqn:p_j_rij0}\\
      s_j(t) &= W_j(p_j^\star(t), t), \label{eqn:sjt_opt}
    \end{align}
  \end{subequations}

  For solutions $(x, \gamma)$ of
  \eqref{eqn:ct_problem}, this implies that $x_j(t) = p_j^\star(t)$
  and $p_i < p_j^\star(t) \implies \gamma_{ij}(t) = 0$.
\end{prop}

The proof can be found in the Appendix,  see \ref{sec:proofs}.

\subsection{Existence and Optimality}
\label{sec:existence_and_optimality}
In this section we address two important technical questions: whether a
solution to our problem actually does exist, and whether the necessary
conditions studied in Section \ref{sec:necessary_conditions} are
sufficient.  Both questions are answered in the affirmative, and
concrete methods for calculating such an optimal solution are provided
in Section \ref{sec:solution_methods}.

\subsubsection{Existence.}
That there exists solutions to the problem \eqref{eqn:ct_problem}
intuitively rests on the assumption that there is a sufficient amount
of supply available to fulfill the contracts.  In the context of our
main application, this is often easily taken for granted due to the
ubiquity of the internet and internet advertising resulting in large
volumes of available impressions.  However, in order to provide an
explicit and interpretable condition, we consider the following
assumption (a version of which also appears in
\cite{leblanc1974transportation}).

\begin{assumption}[Adequate Supply]
  \label{ass:adequate_supply}
  We say that an \textit{adequate supply} condition holds if for every
  $j \in [M]$ we have
  $$\int_0^{\tau_j}B_j(t) \d t > \sum_{i \in \B_j} C_i,$$ where
  $\tau_j = \text{min}\{T_i\ |\ i \in \B_j\}$ and
  $B_j(t) = \underset{x \ge 0}{\text{max }} W(x, t) \le B_W < \infty$.
  \end{assumption}

The above assumption implies that every item type individually has enough supply to
  fulfill each of the contracts to which it's items may be assigned.

With this assumption in hand, we are able to address some important
technical aspects concerning the existence of \textit{regular}
solutions, as well as the smoothness of such solutions.  Recall that
for a solution to be regular means, essentially, that the constraints
are not so stringent as to completely determine the solution.  Our
application in Section \ref{sec:necessary_conditions} of the maximum
principle requires the a-priori knowledge that a regular solution does
in fact exist.  Consult \cite{clarke2013functional} for further detail.

\begin{prop}[Existence]
  \label{prop:existence}
  If there exists a feasible point for Problem \eqref{eqn:ct_problem_cvx},
  then it admits a solution $(s(t), r(t))$.  Moreover, if Assumption
  \ref{ass:adequate_supply} holds, then there exists a feasible point
  and any solution is regular.%   If in addition, the functions
  % $f(t, x), W(t, x)$ are (Lipschitz) continuous in $t$, then there
  % exists a solution which is also (Lipschitz) continuous.
\end{prop}

\noindent{\bf Proof:} See Appendix,  see \ref{sec:proofs}.

\vspace{0.3cm}

\begin{remark}
  A solution $(x, \gamma)$ to \eqref{eqn:ct_problem} can be obtained
  from $(s(t), r(t))$ via
  $$\forall i \in [N]\ x_{ij}(t) = W_j^{-1}(s_j(t), t),$$
  $\gamma_{ij}(t) = r_{ij}(t) / s_j(t)$ where $0/0 = 0$ by convention.
  This is clear from transformations applied to obtain
  \eqref{eqn:ct_problem_cvx} from \eqref{eqn:ct_problem}.  This solution
  inherits the regularity and normality of $(s(t), r(t))$.
\end{remark}

\subsubsection{Optimality.}
The maximum principle we applied in Section
\ref{sec:necessary_conditions} is in essence a manifestation of
Fermat's rule: if $x^\star$ minimizes the smooth function $f$ we must
necessarily have $\nabla f(x^\star) = 0.$ If it is known that the
function $f$ is convex, then this condition is also sufficient, and
any stationary point is a \textit{global} minimum.  The following
proposition (a corollary of \cite[Theorem~24.1]{clarke2013functional})
asserts the analogous result for our problem.

\begin{prop}[Global Optimality]
  \label{prop:global_optimality}
  If a regular solution exists (a sufficient condition being
  Assumption \ref{ass:adequate_supply}), then any pair $(s, r)$
  satisfying the necessary conditions of Proposition
  \ref{prop:optimal_allocation} is globally optimal for Problem
  \eqref{eqn:ct_problem_cvx}.  By extension, the pair $(x, \gamma)$
  derived from $(s, r)$ is globally optimal for
  \eqref{eqn:ct_problem}.
\end{prop}

\noindent{\bf Proof:} 
  This is a corollary of \cite[Theorem~24.1]{clarke2013functional} and
  \ref{prop:existence} since the Hamiltonian does not depend on the
  state and since the objective function and constraint region is
  convex.

\section{Solution Methods}
\label{sec:solution_methods}
The Problem \eqref{eqn:ct_problem} can be reformulated as a convex
problem, but with an uncountable infinite number of variables.  In
this section, we establish the fact that a \textit{piecewise constant}
solution exists, and therefore that the entire problem can be reduced
into a finite dimensional optimization problem, and again formulated
as a finite convex problem and solved by well known methods.  We focus
back on Problem \eqref{eqn:ct_problem} because the upcoming
Proposition \ref{prop:piecewise_constant_allocation} is easier to
state and to understand than the equivalent statement for
\eqref{eqn:ct_problem_cvx}.

Combining the results of Propositions \ref{prop:ubp} and
\ref{prop:optimal_allocation}, as well as $f_j(0, t) = 0$, we can
narrow down the properties of the optimal solution, and reformulate
Problem \eqref{eqn:ct_problem} as
\begin{equation}
  \label{eqn:ct_problem_p}
  \begin{aligned}
    \underset{p, \gamma, q}{\text{minimize}} \quad & \sum_{j = 1}^M \int_0^{T^j} f_j(q_j(t), t) \d t\\
    \textrm{subject to}
    \quad & \int_0^{T_i}\Big[\sum_{j \in \A_i}\gamma_{ij}(t)W_j(q_j(t), t)\Big]\d t \ge C_i\\
    \quad & \sum_{i \in \B_j \cap \Tt} \gamma_{ij}(t) = \ind[t < T^j]\\
    \quad & i \in \B_j, p_i < q_j(t) \implies \gamma_{ij}(t) = 0,\\
    \quad & q_j(t) = \underset{i \in B_j \cap \Tt}{\text{max}} p_i, p_i \ge 0, \gamma_{ij}(t) \ge 0.
  \end{aligned}
\end{equation}

It is convenient introduce the discrete analog of the set $\Tt$,
namely, $$\Tk \defeq \{i\; |\; T_i \le T_k\}.$$ We are now able to
establish the existence of a piecewise constant solution.

\begin{prop}[Piecewise Constant Allocation]
  \label{prop:piecewise_constant_allocation}
  There exists piecewise constant functions $q(t), \gamma(t)$ taking values
  $q_{j}[k], \gamma_{ij}[k]$ for times $t \in [T_{k - 1}, T_k)$ which are optimal for
  Problem \eqref{eqn:ct_problem_p}.
\end{prop}

\noindent{\bf Proof:}
  
  Let $(p, q, \gamma)$ be a solution to \eqref{eqn:ct_problem_p}.

  First, any $q$ solving \eqref{eqn:ct_problem_p} is already
  piecewise constant by the definition of $\Tt$, and the constraint
  $q_j(t) = \underset{i \in \B_j \cap \Tt}{\text{max}}\; p_i$.

  Since the objective does not depend on $\gamma(t)$, we only need to
  find a feasible piecewise constant constant $\gamma(t)$.  To do so,
  define
  $$H_{ij}(k) \defeq \int_{T_{k - 1}}^{T_k}\gamma_{ij}(t)W_j(q_j(t), t) \d t.$$
  Then, since $\gamma(t)$ forms part of a solution, for each
  $i \in [N]$ we have
  $$\sum_{j \in \A_i}\sum_{k: T_k \le T_i} H_{ij}(k) \ge C_i.$$  Now, for $k: T_k \le T^j$ let
  $$H_j(k) \defeq \sum_{i \in \B_j}H_{ij}(k) \overset{(a)}{=} \int_{T_{k - 1}}^{T_k}W_j(q_j(t), t)\d t,$$ where the latter equality follows since $\sum_{i \in \B_j \cap \Tt} \gamma_{ij}(t) = 1; \forall t:\; \B_j \cap \Tt \ne \emptyset.$

  Define now the piecewise constant allocation
  $$t \in [T_{k - 1}, T_k), i \in \Tt \implies \tilde{\gamma}_{ij}(t) = \gamma_{ij}[k]\defeq \frac{H_{ij}(k)}{H_j(k)}\ind[T_k \le T^j].$$  We see that this function is feasible firstly since $\tilde{\gamma}_{ij}(t) \ge 0$ and $\sum_{i \in \B_j \cap \Tt}\gamma_{ij}[k] = \ind[T_k \le T^j]$ by construction.  Moreover, $i \in \B_j\; p_i < q_j(t) \implies \tilde{\gamma}_{ij}(t) = 0$ is a property inherited from $\gamma(t)$ by definition of $H_j, H_{ij}$.  Finally $$\sum_{j \in \A_i}\sum_{k:T_k \le T_i}\gamma_{ij}[k]\int_{T_{k - 1}}^{T_k}W_j(q_j, t) \d t = \sum_{j \in \A_i}\sum_{k:T_k \le T_i}H_{ij}(k) \ge C_i.$$

  by $(a)$ above.
\vspace{0.3cm}

By incorporating the results of Proposition
\ref{prop:piecewise_constant_allocation}, Problem
\eqref{eqn:ct_problem_p} can be written as a finite optimization
problem.  However, the final two constraints are, as written, still
problematic.  Our main theorem shows that these constraints can simply
be dropped, and serves to summarize our developments by establishing
that solutions of the resulting problem can be converted into globally
optimal solutions of the original optimal control problem
\eqref{eqn:ct_problem}.  It will be seen that the general problem
considered in this paper, can in fact be reduced exactly to an
instance of the seemingly less general static problem considered in
\cite{marbach_bidding_2020}.

\begin{thm}[Optimal Solution]
  \label{thm:maintheorem}
  Let $\fjk(x) = \int_{T_{k - 1}}^{T_k}f_j(x, t) \d t$,
  $\Wjk(x) = \int_{T_{k - 1}}^{T_k}W_j(x, t) \d t$, and
  $\Ljk = \fjk \circ \Wjk^{-1}$.  Consider the following optimization
  problem
  \begin{equation}
    \label{eqn:problem_finite}
    \begin{aligned}
      \underset{s, r}{\mathrm{minimize}} \quad & \sum_{j = 1}^M\sum_{k: T_k \le T^j} \Ljk(s_j[k])\\
      \mathrm{subject\; to}
      \quad & \sum_{j \in \A_i}\sum_{k: T_k \le T_i} r_{ij}[k] \ge C_i\\
      \quad & \sum_{i \in \B_j \cap \Tk} r_{ij}[k] = s_j[k]\\
      \quad & r_{ij}[k] \ge 0.
    \end{aligned}
    \tag{$P^\star$}
  \end{equation}

  This problem is convex, and if it's solutions $s_j[k], r_{ij}[k]$ are
  transformed into functions $s_j(t), r_{ij}(t)$ of $t \in [0, T]$
  according to $s_j(t) = s_j[k]$ and $r_{ij}(t) = r_{ij}[k]$ if
  $t \in [T_{k - 1}, T_k)$ results in a globally optimal solution to
  Problem \eqref{eqn:ct_problem_cvx}.

  Similarly, by transforming
  $\gamma_{ij}[k] \defeq r_{ij}[k] / s_j[k]$ and
  $x_j[k] \defeq \Wjk^{-1}(s_j[k])$ into continuous functions results
  in a globally optimal solution to Problem \eqref{eqn:ct_problem}.
\end{thm}

\noindent{\bf Proof:} The details can be found in the Appendix,  see \ref{sec:proofs}.

\begin{remark}[Implementation]
  The Problem \eqref{eqn:problem_finite} is equivalent to the
  time-homogeneous version of the problem studied by
  \cite{marbach_bidding_2020} with compound item types $(j, k)$ and
  supply curves $\Wjk$.  The valid (compound) types for contract $i$:
  $\overline{\A}_i \defeq \A_i \times \{k\; |\; T_k \le T_i \}$ and
  set of valid contracts for type $(j, k)$:
  $\overline{\B}_{jk} \defeq \B_j \cap \Tk$. Hence one can use the
  algorithm in \cite{marbach_bidding_2020} that does not involve
  derivatives of $W_j$. Problem \eqref{eqn:problem_finite} can also be
  solved by standard convex optimization software. Indeed, we used
  CVXOPT \cite{vandenberghe2010cvxopt} for the numerical results in
  Section \ref{sec:simulations}.

  In practice, if the problem \eqref{eqn:problem_finite} is infeasible
  (i.e., the supply is not adequate c.f. Assumption
  \ref{ass:adequate_supply}), a ``best-effort'' set of bids can be
  computed by instead using a penalty formulation, e.g. with cost
  function
  \begin{equation*}
      \sum_{j = 1}^M\sum_{k: T_k \le T^j} \Ljk(s_j[k]) + \rho\Bigl[\sum_{i = 1}^NC_i - \sum_{j \in \A_i}\sum_{k: T_k \le T_i} r_{ij}[k]\Bigr],
  \end{equation*}
  and a very large value of $\rho$.
\end{remark}

\section{Simulation}
\label{sec:simulations}
To evaluate and illustrate the performance of algorithms derived in
this paper we carried out a set of numerical simulations on data
derived from the iPinYou dataset \cite{liao2014ipinyou,
  zhang2014real}.  All of the computations have been carried out with
Python's scientific computing ecosystem \cite{2020SciPy-NMeth}.

Using the top $5$ highest rate item types from the IPinYou dataset
(i.e., $j \in [5]$), we  constructed $6$ contracts with time
deadlines, supply requirements, and targeting sets according to Table
\ref{tab:contract_table}.  Without loss of generality we assumed that all contracts began at $t = 0$.

\begin{table}
  \centering
  \captionsetup{type=table}
  \caption{Simulation Contract Specifications}
  \label{tab:contract_table}
  \begin{tabular}{cccc}
    \toprule
    $i$ &  $T_i$ (hours) & $C_i$ & $\A_i$ \\
    \midrule
    $1$ & $28$ & $4500$ & $\{0, 2\}$ \\
    $2$ & $31$ & $3240$ & $\{0, 4\}$ \\
    $3$ & $43$ & $6300$ & $\{1, 2, 4\}$ \\
    $4$ & $56$ & $3600$ & $\{0, 3\}$ \\
    $5$ & $63$ & $1800$ & $\{2\}$ \\
    $6$ & $71$ & $3600$ & $\{2, 4\}$ \\
    \bottomrule
  \end{tabular}
\end{table}

A sliding window was used in the simulations for the one week of data 
available: each simulation period spanned a 72 hour
long window beginning every 12 hours.  Thus, a simulation was run on
hours 0 through 72, 12 through 84, 24 through 96, etc.  The purpose of
the sliding window is to capture the variance in bid paths resulting
from day to day changes in market conditions.  We additionally repeated
each simulation 4 times in order to capture the variance arising from
the randomness inherent in the bidding strategy.  There are a total of
9 unique periods and therefore 36 simulations per algorithm in total.

To facilitate interpretation and plotting, we 
re-normalize the simulation results by scaling time and supply
requirements.  If $c_i(t)$ is the total supply attained for contract
$i$ by time $t$, we define $\tilde{c}_i(t) \defeq c_i(tT_i) / C_i$ so
that $\tilde{c}_i(1) \ge 1$ indicates that contract $i$ has been
fulfilled by it's deadline.  Finally, we can average each of these
curves into a single function
$\tilde{c}(t) \defeq \frac{1}{N}\sum_{i = 1}^N \tilde{c}_i(t)$ so that
$\tilde{c}(1) \ge 1$ indicates that \textit{every} contract has been
fulfilled by their deadlines.  Our figures depict this averaged and
re-normalized curve.

\begin{figure}
  \centering
  \caption{Contract Management Discrete Event Simulation}
  \label{fig:simulation_results}

  \begin{subfigure}[b]{0.45\textwidth}
    \caption{\centering Normalized Acquisition Paths $\tilde{c}(t)$}
    \label{fig:oos_simulation_main}
    \includegraphics[width=\textwidth]{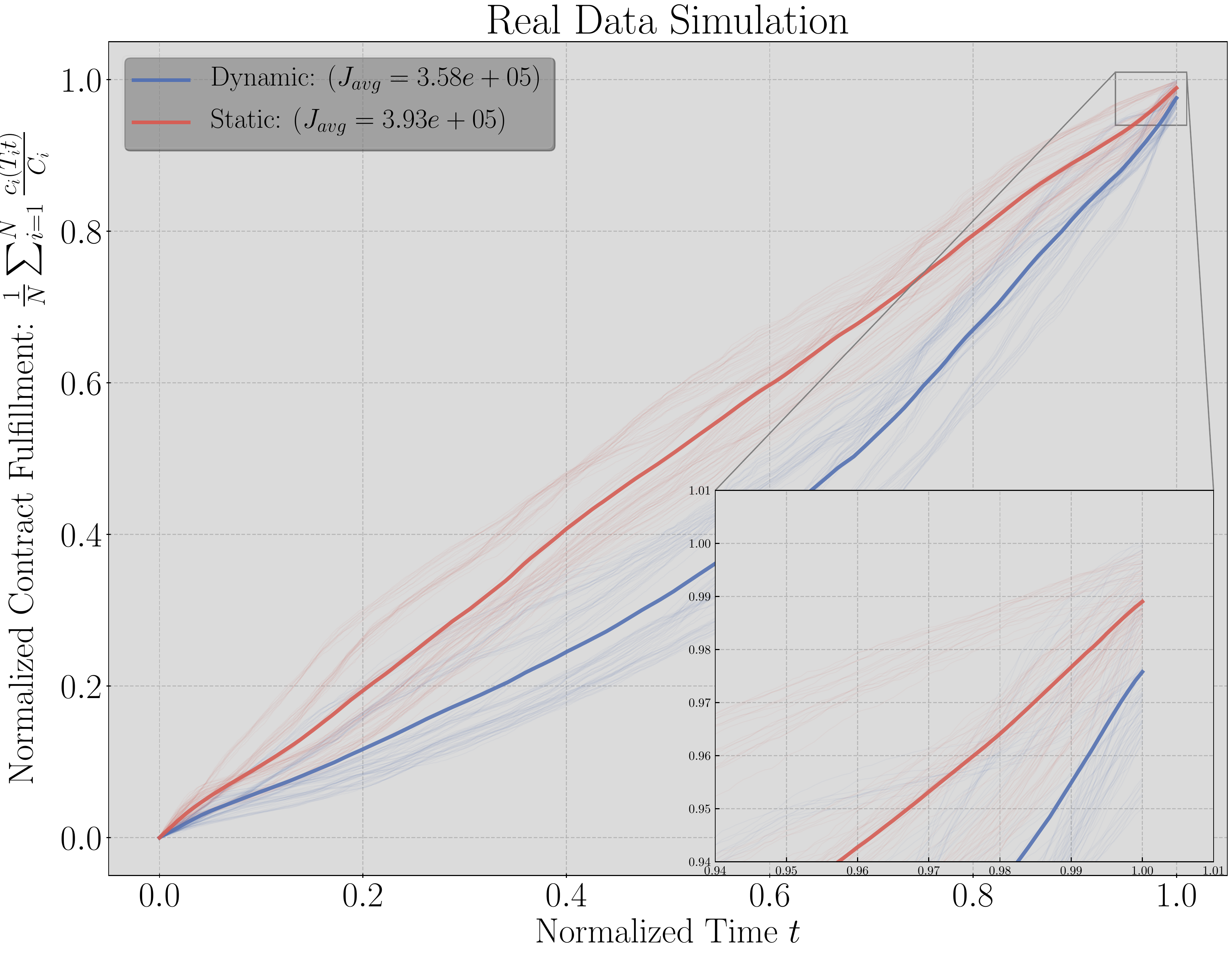}
  \end{subfigure}
  \begin{subfigure}[b]{0.45\textwidth}
    \caption{\centering Bid Paths $\rho(t)$}
    \label{fig:bid_paths}
    \includegraphics[width=\textwidth]{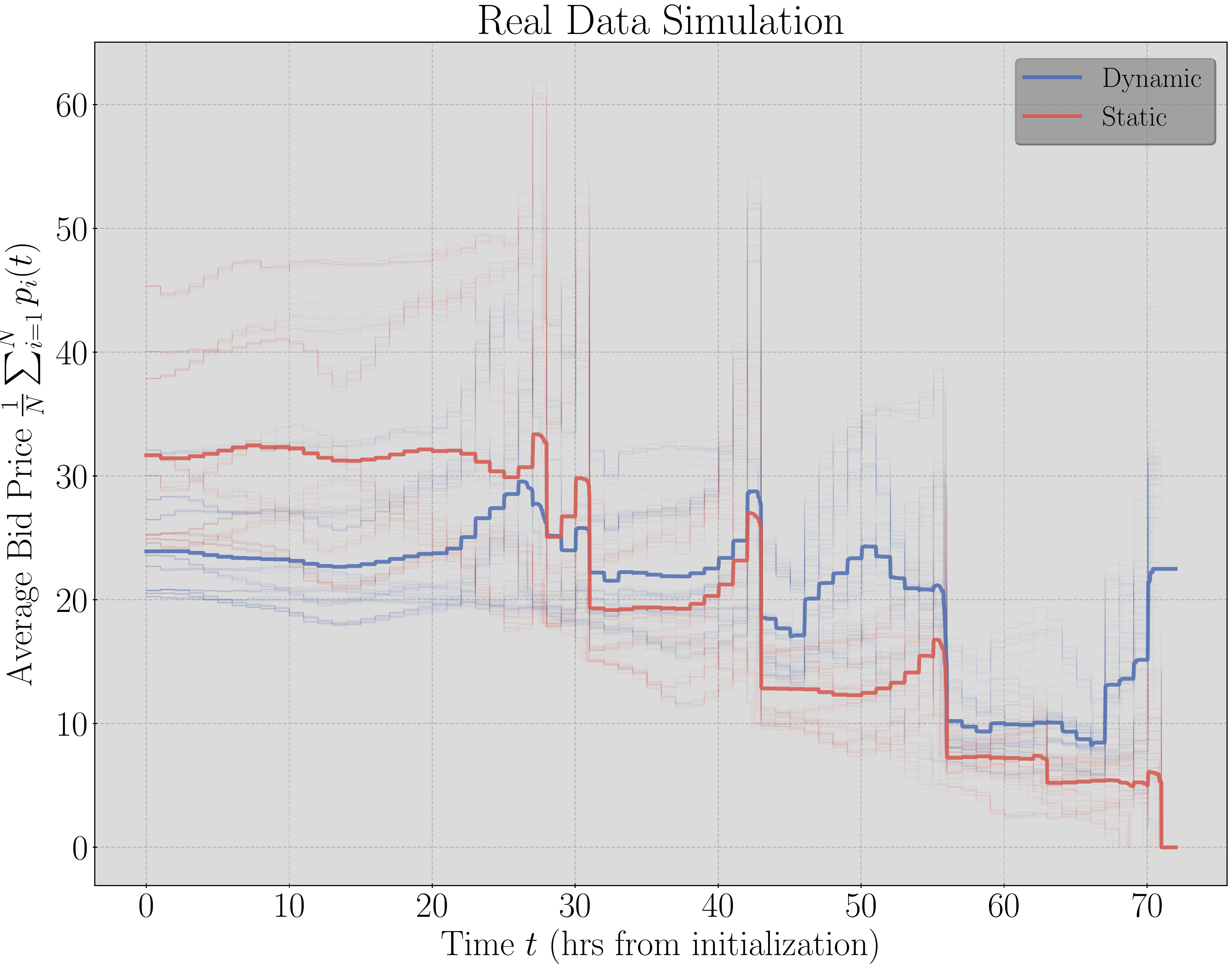}
  \end{subfigure}

  \footnotesize{Discrete event simulations with IPinYou data.  Thin
    and lightly shaded lines depict a single simulation with the thick
    dark line being the mean.  $(a)$ Plots of the averaged and
    re-normalized acquisition paths $\tilde{c}(t)$ for the contracts
    described in Table \ref{tab:contract_table}.  Bids for the
    \textit{Dynamic} solution (blue) are calculated according to
    Problem \ref{eqn:problem_finite}, and the \textit{Static} solution
    (red) is calculated similarly, except averages are taken over the
    entire $[0, T]$ period.  In both cases, a receding horizon of one
    hour is employed to update bids over time.  The average cost
    across simulations is denoted $J_{avg}$ (in the legend) and is in
    abstract currency units.  $(b)$ The average (across contracts) bid
    path $p(t)$ corresponding to the simulations of $(a)$.  Large
    discontinuities correspond to contract fulfillment times, and
    small adjustments to the hourly receding horizon.}
\end{figure}

We compared the results using \ref{thm:maintheorem} referred to as the
{\em dynamic solution} with a {\em static} approach based on averaging
supply and impression constraints over the contract duration. The
supply curve for the averaged system is taken to be
$\overline{W}_j(x) = \frac{1}{T}\int_0^TW_j(x, t)\mathsf{d} t$. Such a
time homogenous solution is a natural heuristic and provides a
baseline.

The whole dataset was used to estimate the supply curves as
24hr-periodic functions by taking the product of a Gaussian kernel
density estimate of the price (for each hour) and the average arrival
rate by hour.  This was extended via periodicity to the entire
week-long period. In practice, simple average prices and arrival rates
could be estimated from historical data available to any DSP. The
estimated supply curves were then used for bidding with the impression
arrivals sampled from the dataset.  Details are provided in Appendix
\ref{sec:estimating_supply_curves}.

Figure \ref{fig:simulation_results} provides a comparison between
these two approaches. For both algorithms, we recompute a new bid
every 1 hour of simulated time.  In the case of the the dynamic
solution, updating the bids at time $\tau \in [0, T)$ requires
re-aggregating active (i.e.  $\tau \in [T_{k - 1}, T_k)$) supply curves via
$\overline{W}_{jk}(x) = \int_\tau^{T_k}W_j(x, t)\d t$ before recomputing new
bids, c.f. Theorem \ref{thm:maintheorem}.  Figure
\ref{fig:bid_paths} depicts the average bid across campaigns
$\frac{1}{N}\sum_{i = 1}^N p_i(t)$ with the time axis being real time and
where the hourly bid updates are clearly discernible.

The results demonstrate the benefits conferred by accounting for time
dynamics and different durations: the simulation results for the
dynamic and static algorithms are $3.58 \times 10^5$ and
$3.93 \times 10^5$ respectively, an average improvement for the
dynamic policy of about $10\%$ for the iPinYou dataset.

\section{Conclusion}
\label{sec:conclusion}
This paper has studied a control problem faced by a DSP obligated to
acquire, on the RTB market, a certain number of items by a given time
deadline.  We have shown that the notion of a supply curve emerges
naturally from the market dynamics and that the optimal contract
management problem can be formulated in terms of these supply curves,
which act essentially as \textit{information states} for the DSP.

Using the Pontryagin maximum principle as our primary tool, we have
analyzed how the structure of the optimal bids depends upon a
targeting set decomposition, the set of campaigns, and the time
deadlines.  We used these results to derive a globally optimal bidding
algorithm from the solution of a convex optimization problem.  The
resulting optimization problem turns out to be a generalization of the
simpler 1-period problem of \cite{marbach_bidding_2020} and the
Transportation-Production problems of \cite{leblanc1974transportation,
  sharp1970decomposition}.

Our algorithm has been illustrated through simulation with real
auction data, demonstrating the potential for improvement over and
above a strictly average case time-homogeneous method.

\clearpage
\printbibliography
\clearpage

\appendix
\section{Proofs}
\label{sec:proofs}

In this section we present the proofs of technical results in the paper.

\noindent{\bf Proof of Proposition \ref{prop:supply_curve_properties}}

  That the formula for $W_j^\bM$ is a CDF is clear since it is monotone
  non-decreasing, right-continuous, and
  $W_j(x, t) \rightarrow 1 \text{ as } x \rightarrow \infty$.

  Now when an item arrives, since each of the participants choose
  independently with probabilities $r_{ij}$ whether or not to bid on
  the item, the probability that bidder $0$ wins the item with bid
  $x \ge 0$ is given by the probability that no one with a bid greater
  than $x$ decides to bid (we use $\prod_{k \in \emptyset} \bullet \defeq 1$):

  \begin{equation*}
    W_j^\bM(x, t) = \prod_{k: b_{kj} > x} (1 - r_{kj}).
  \end{equation*}

  The result then follows by simple algebra:

  \begin{align*}
    \prod_{k: b_{kj} > x} (1 - r_{kj})
    &= \prod_{k: b_{kj} > x} \textup{exp}(\ln(1 - r_{kj}))\\
    &= \textup{exp}\Big(-\sum_{k: b_{kj} > x} \phi(r_{kj})\Big)\\
    &= \textup{exp}\Big(-\sum_{i = 1}^N \phi(r_{ij})\ind_{(x, \infty)}(b_{ij})\Big),
  \end{align*}

  where we define $W_j^\bM(x) = 0$ for any $x < 0$.
\vspace{0.5cm}

\noindent{\bf Proof of Proposition \ref{prop:expected_win_rate}:}

  Consider the summation
  $$\sum_{i = 1}^{N(t)} \phi(r_i)\ind_{(x, \infty)}(b_i).$$
  Firstly, since $\ind_{(x, \infty)}(b_i) \sim \text{Ber}(1 - F_B(x))$
  the number of non-zero elements in the summation is
  $\text{Bin}(N(t), 1 - F_B(x))$ distributed.  Then, the Poisson
  thinning property implies that the number of terms in the summation
  remains Poisson with parameter $\rho_B(x)$.

  Now, following from Proposition~\ref{prop:supply_curve_properties},
  using the independence of $N(t), r_i, b_i$, and conditioning
  on the number of summation terms, we obtain:

  \begin{align*}
    \E[W^\bM(x, t)]
    &= \E\textup{exp}\Bigl(-\sum_{i = 1}^{N_B} \Phi(r_i)\ind_{(x, \infty)}(b_i)\Bigr)\\
    &= \sum_{n = 0}^\infty \frac{e^{-\rho_B(x)}\rho_B(x)^n}{n!} \E\textup{exp}\Bigl(-\sum_{i = 1}^n \Phi(r_i)\Bigr).\\
    &= \sum_{n=0}^{\infty}  \frac{e^{-\rho_B(x)}\rho_B(x)^n}{n!} \prod_{i=1}^n \E[e^{-\Phi(r_i)}]\\
    &= \sum_{n=0}^{\infty}  \frac{e^{-\rho_B(x)}\rho_B(x)^n}{n!} \prod_{i=1}^n (1-\E[r_i])\\
    &= e^{-\rho_B(x)\E[r] }\sum_{n=0}^{\infty}  \frac{e^{-\rho_B(x)(1-\E[r])}(\rho_B(x)(1-\E[r]))^n}{n!} \\
    &= e^{-\rho_B(x) \E[r]}
  \end{align*}
  
  Noting that $\rho_B(x) = \rho(1-F_B(x)) $ we see that $W(x)$ is a
  right-continuous increasing function and $W(x) \rightarrow 1$ as
  $x \rightarrow \infty$, hence it is a distribution.

\vspace{0.5cm}

\noindent{\bf Proof of Proposition \ref{prop:smoothed_probabilities}}

  Fix some $\epsilon > 0$, and compact set $I \subset \R$, as well as
  a parameter $\delta > 0$.  Then, denoting
  $\phi_\sigma(x) \defeq
  \frac{e^{-\frac{x^2}{2\sigma^2}}}{\sqrt{2\pi\sigma^2}}$ the standard
  Gaussian density and
  $\mu_\alpha \defeq \E[|\mathcal{X} - \E[\mathcal{X}]|^\alpha]$ the
  centered $\alpha$-moment of a standard Gaussian\footnote{precisely,
    $\mu_\alpha = 2^{\alpha/2}\Gamma(\frac{\alpha + 1}{2})/\sqrt{\pi}$
    according to \cite{winkelbauer2012moments}.}.  We then have,
  \begin{align*}
    \int_I |\Ws(x) - W(x)| \d x
    &= \int_I\Big| \int_\R W(t)\phi_\sigma(x - t)\d t - W(x) \Big| \d x\\
    &\overset{(a)}= \int_I\Big| \int_\R \big(W(t) - W(x)\big)\phi_\sigma(x - t)\d t\Big| \d x\\
    &\overset{(b)}= \sum_{k = 1}^n \int_{I \cap B(x_k; \delta)} \Big|\int_\R \big(W(t) - W(x)\big)\phi_\sigma(x - t)\d t\Big| \d x, \\
    & \qquad + \int_{I \setminus \bigcup_{k = 1}^n B(x_k; \delta)} \Big|\int_\R \big(W(t) - W(x)\big)\phi_\sigma(x - t)\d t \Big| \d x,\\
  \end{align*}

  where $(a)$ follows since $\int_\R \phi_\sigma(x)\d x = 1$, and
  $(b)$ by breaking up the first integral with balls $B(x_k, \delta)$
  at the jump points $x_1, \ldots, x_n$.  Then,
  \begin{align*}
    \int_I |\Ws(x) - W(x)| \d x
    &\le \sum_{k = 1}^n \int_{I \cap B(x_k; \delta)} \int_\R \big|W(t) - W(x)\big|\phi_\sigma(x - t)\d t\d x \\
    & \qquad + \int_{I \setminus \bigcup_{k = 1}^n B(x_k; \delta)} \int_\R \big|W(t) - W(x)\big|\phi_\sigma(x - t)\d t \d x\\
    &\overset{(a)}\le \sum_{k = 1}^n \int_{I \cap B(x_k; \delta)} \int_\R \big(L|t - x|^\alpha + B\big)\phi_\sigma(x - t)\d t\d x\\
    & \qquad + \int_I \int_\R L|t - x|^\alpha\phi_\sigma(x - t)\d t \d x,\\
  \end{align*}

  where $(a)$ follows via the definition of $(L, \alpha)$-H\"older
  continuity.  We then evaluate the centered moments to obtain:
  \begin{align*}
    \int_I |\Ws(x) - W(x)| \d x
    &\le \sum_{k = 1}^n \int_{I \cap B(x_k; \delta)} (L\mu_\alpha\sigma^\alpha + B) \d x + \int_I L\mu_\alpha\sigma^\alpha \d x\\
    &\le nB\delta + (n\delta + |I|) L\mu_\alpha\sigma^\alpha.\\
  \end{align*}

  Finally, let $\sigma = \delta^{1/2}$ and take the limit:
  \begin{align*}
    nB\delta + (n\delta + |I|) L\mu_\alpha\sigma^\alpha
    &= nB\delta + nL\mu_\alpha\delta^{\frac{\alpha + 2}{2}} + |I|L\mu_\alpha \delta^{1/2}\\
    &\rightarrow 0 \text{ as } \delta \rightarrow 0.
  \end{align*}

  When there are no jumps, (or at regions outside the jumps) we
  similarly have a uniform approximation simply by applying H\"older
  continuity and the triangle inequality.

  That $\Ws$ is $C_{\infty}$ simply follows by Leibniz's integral
  rule, the smoothness of $e^{-x}$, and boundedness of $W$:
  \begin{align*}
    \frac{\d\Ws}{\d x}(x)
    &= \frac{1}{\sqrt{2\pi\sigma^2}}\int_{-\infty}^\infty W(t)\frac{\d}{\d x}\Big[\text{exp}\big(-\frac{(t - x)^2}{2\sigma^2}\big)\Big]\d t\\
    &= \frac{1}{\sigma^2\sqrt{2\pi}}\int_{-\infty}^\infty W(t)(t - x)\text{exp}\big(-\frac{(t - x)^2}{2\sigma^2}\big)\d t.\\
  \end{align*}

  Finally, we see from here that $\Ws'(x) > 0$ since for
  any $x$ s.t. $W(x) > 0$ we have
  \begin{align*}
    \Ws'(x)
    &= \frac{1}{\sigma^2\sqrt{2\pi}}\int_{0}^\infty W(t)(t - x)\phi_\sigma(t - x)\d t\\
    &= \frac{1}{\sigma^2\sqrt{2\pi}}\int_{0}^x W(t)(t - x)\phi_\sigma(t - x)\d t + \frac{1}{\sigma^2\sqrt{2\pi}}\int_{x}^\infty W(t)(t - x)\phi_\sigma(t - x)\d t\\
    &\overset{(a)}{\ge} \frac{1}{\sigma^2\sqrt{2\pi}}\int_{0}^x W(x)(t - x)\phi_\sigma(t - x)\d t + \frac{1}{\sigma^2\sqrt{2\pi}}\int_{x}^\infty W(x)(t - x)\phi_\sigma(t - x)\d t\\
    &= \frac{W(x)}{\sigma^2\sqrt{2\pi}}\int_{0}^{\infty}(t - x)\phi_\sigma(t - x)\d t\\
    &> 0,
  \end{align*}

  where $(a)$ follows from the (weak-) monotonicity of $W$, and
  noticing that $t - x < 0$ in the first integral and $t - x > 0$ in
  the second.  If $W(x) = 0$ we can replace the inequality in $(a)$
  with

  $$
  \ge \frac{1}{\sigma^2\sqrt{2\pi}}\int_{x}^\infty W(t)(t - x)\phi_\sigma(t - x)\d t > 0,\\
  $$

  and therefore $\Ws$ is strictly monotone increasing.

  The final statement that
  $x\Ws(x) \rightarrow 0 \text{ as } x \rightarrow -\infty$ follows
  since for $x < 0$ we have $W(x) = 0$ and the Mills' ratio (ratio of
  the complementary c.d.f. to the p.d.f.) of $\mathcal{N}(0, 1)$ is
  asymptotically $1/x$.

\vspace{0.5cm}

\noindent{\bf Proof of Proposition \ref{prop:optimal_allocation}}

  The statement concerning $c_i(T_i) \ge C_i$ and $p_i \ge 0$ has
  already been established in the main text.

  Now, look at a fixed time $t$, let $s$ form part of an optimal
  solution to Problem \eqref{eqn:sc_hamiltonian_problem}.  We consider
  optimization over $r$ alone as
  \begin{equation}
    \begin{aligned}
      \underset{r}{\text{maximize}} \quad & \sum_{i \in \Tt}\sum_{j \in \A_i} p_ir_{ij} - \sum_{j \in \Ttj}\Lambda_j(s_j, t)\\
      \textrm{subject to}
      \quad & \sum_{i \in \B_j \cap \Tt}r_{ij} = s_j, r_{ij} \ge 0.\\
    \end{aligned}
  \end{equation}

  We first swap the order of summation in the objective:
  \begin{equation*}
    \sum_{i \in \Tt}\sum_{j \in \A_i} p_ir_{ij} = \sum_{j \in \Ttj} \sum_{i \in \B_j \cap \Tt} p_i r_{ij},
  \end{equation*}

  which is valid since
  $(i, j) \in \Tt \times \A_i \iff (j, i) \in \Ttj \times (\B_j \cap
  \Tt).$ To see this, note that $j \in \A_i \iff i \in \B_j$.  and if
  $i \in \Tt$, then for any $j \in \A_i$ we necessarily have $t < T^j$
  and therefore $j \in \Ttj$.

  Applying H\"older's inequality to the second summation and using the
  problem constraints we have the inequality
  \begin{equation*}
    \sum_{j \in \Ttj} \sum_{i \in \B_j \cap \Tt} p_i r_{ij} \le \sum_{j \in \Tt} p_j^\star(t) s_j(t).
  \end{equation*}

  If Condition \eqref{eqn:p_j_rij0} is satisfied, then this value is
  achieved.

  Now, suppose that $r_{ij} > 0$, but $p_i < p_j^\star(t)$.  Then, the
  objective value can be increased by reassigning
  $r_{ij} \leftarrow 0$ and
  $r_{i^\star j} \leftarrow r_{i^\star j} + r_{ij}$, where $i^\star$
  is such that $p_{i^\star} = p_j^\star(t)$.  If $p_j^\star(t) = 0$,
  the statement is vacuous.  Therefore, any $r$ which fails to satisfy
  \eqref{eqn:p_j_rij0} cannot be a solution.

  We turn to the Lagrangian necessary conditions (see
  e.g. \cite[Chap.~9]{clarke2013functional} or
  \cite[Chap.~3]{bertsekas1997nonlinear}) for Problem
  \eqref{eqn:sc_hamiltonian_problem}.  We consider the Lagrangian of
  of the equivalent minimization problem (swapping the sign of the
  objective, since multiplier theorems are typically stated with this
  convention) to obtain
  \begin{equation}
    \L(s, r, \mu, \theta) = -\sum_{i \in \Tt}\sum_{j \in \A_i}(p_i + \theta_{ij}) r_{ij} + \sum_{j \in \Ttj}\Bigl[\mu_j \Bigl(\sum_{i \in \B_j \cap \Tt} r_{ij} - s_j \Bigr) + \Lambda_j(s_j, t)\Bigr],
  \end{equation}

  and it's derivative, where $(i, j)$ satisfies $i \in \B_j \cap \Tt, j \in \A_i$:
  \begin{equation}
    \begin{aligned}
      \frac{\partial\L}{\partial r_{ij}} &= -p_i - \theta_{ij} + \mu_j\\
      \frac{\partial\L}{\partial s_j} &= -\mu_j + \Lambda_j'(s_j, t).
    \end{aligned}
  \end{equation}

  We consider the first order condition for $(x, \gamma)$ to be a
  solution to \eqref{eqn:sc_hamiltonian_problem} and seek multipliers
  $(\mu, \theta)$ such that $\theta_{ij} \ge 0$,
  $\theta_{ij}r_{ij} = 0$, and
  $\frac{\partial\L}{\partial \gamma_j}(s, r, \mu, \theta) =
  \frac{\partial\L}{\partial x_j}(s, r, \mu, \theta) = 0.$ Due to the
  convexity of the objective and the linearity of the constraints, the
  existence of such multipliers is necessary and sufficient for
  optimality.  We therefore consider the following:
  \begin{subequations}
    \label{eqn:ham_lag}
    \begin{align}
      \mu_j = \theta_{ij} + p_i, \label{eqn:ham_lag1}\\
      \Lambda_j'(s_j, t) = \mu_j, \label{eqn:ham_lag2}
    \end{align}
  \end{subequations}

  and make the ansatz $\theta_{ij} = p_j^\star(t) - p_i$,
  $\mu_j = p_j^\star(t)$.  This choice of $\theta$ satisfies
  $\theta_{ij} \ge 0$ (by definition) and by our earlier assertion
  \eqref{eqn:p_j_rij0} we have complementary slackness:
  $\theta_{ij}r_{ij} = 0$.

  If $p_j^\star(t) > 0$ then $s_j = \Lambda_j'^{-1}(p_j^\star(t))$.  And if
  $p_j^\star(t) = 0,$ we may take $s_j(t) = W_j(0),$ (see Proposition
  \ref{prop:convex_acquisition_costs}) which is consistent.  Equation
  \eqref{eqn:sjt_opt} follows, and the final statement is immediate by
  the definition of the transformation between $(x, \gamma)$ and
  $(s, r)$.
  
\vspace{0.3cm}

\noindent{\bf Proof of Proposition \ref{prop:existence}}

  If we augment Problem \eqref{eqn:ct_problem_cvx} with the constraint
  $s_j(t) \le B_j(t)$, then the existence of a solution follows from
  \cite[Theorem~23.11]{clarke2013functional} using convexity and the
  boundedness of the control set, as well as the (assumed) existence
  of a feasible point.

  Supposing henceforth that Assumption \ref{ass:adequate_supply}
  holds, in this case the existence of a feasible point is evident.
  Suppose by way of contradiction that a solution $(s(t), r(t))$ is
  non-regular solution with corresponding adjoint $p \in \R_+^N$ (see
  Section \ref{sec:necessary_conditions}).  Then, at a particular
  point in time $\tau$, where $s, r = s(\tau), r(\tau)$ maximizes
  $\sum_{i = 1}^N p_i \sum_{j \in \A_i} r_{ij}$ (i.e., the solution is
  independent of the cost function), subject to the constraint
  $\sum_{i \in \B_j}r_{ij} = s_j$, which is independent of the cost.
  Since there must be at least one $p_i > 0$ (otherwise we would
  violate the non-triviality condition
  \cite[Theorem~22.26]{clarke2013functional}) we must have
  $s_j(t) = B_j(t)$ for at least one $j \in \A_i$.  However, by the
  assumption, this must oversupply contract $i$ (and others), and
  therefore by the requirement that $-p_i(T) \in N_{E_i}(c_i(T))$ (see
  Equation \eqref{eqn:normal_cone}) we must have $p_i = 0$, which is a
  contradiction.  Therefore, the solution must be regular.

\vspace{0.3cm}

\noindent{\bf Proof of Theorem \ref{thm:maintheorem}:}

  We first establish convexity, and then use Proposition
  \ref{prop:piecewise_constant_allocation} to show that solutions of
  \eqref{eqn:problem_finite} can be converted into solutions of
  \eqref{eqn:ct_problem_p} and therefore to solutions of
  \eqref{eqn:ct_problem_cvx} and \eqref{eqn:ct_problem}.

  Calculate, for $x \ge 0$:
  \begin{align*}
    \fjk(x)
    &= \int_{T_{k - 1}}^{T_k} f_j(x, t)\d t\\
    &= \int_{T_{k - 1}}^{T_k} \int_0^x u W_j'(u, t)\d u\d t\\
    &= \int_0^xu \int_{T_{k - 1}}^{T_k}W_j'(u, t)\d t\d u\\
    &= \int_0^xu \Wjk'(u)\d u,
  \end{align*}

  so $\fjk$ is just the cost function for a second price auction with
  strictly monotone supply curve $\Wjk$.  Therefore, the results of
  Proposition \ref{prop:convex_acquisition_costs} hold for $\Ljk$.

  Now, using Proposition \ref{prop:piecewise_constant_allocation},
  convert Problem \eqref{eqn:ct_problem_p} into the finite
  optimization problem
  \begin{equation}
    \begin{aligned}
      \underset{p, q, \gamma}{\text{minimize}} \quad & \sum_{j = 1}^M\sum_{k: T_k \le T^j} \fjk(q_j[k])\\
      \textrm{subject to}
      \quad & \sum_{j \in \A_i} \sum_{k: T_k \le T_i} \gamma_{ij}[k]\Wjk(q_j[k]) \ge C_i\\
      \quad & \sum_{i \in \B_j \cap \Tk} \gamma_{ij}[k] \le \ind[T_k \le T^j]\\
      \quad & i \in \B_j, p_i < q_j[k] \implies \gamma_{ij}[k] = 0,\\
      \quad & q_j[k] = \underset{i \in B_j \cap \Tk}{\text{max}} p_i, p_i \ge 0, \gamma_{ij}[k] \ge 0.
    \end{aligned}
  \end{equation}

  Applying similar transformations as in Section
  \ref{sec:convex_reformulation}, this is equivalent to
  \begin{equation}
    \label{eqn:finite_problem_full}
    \begin{aligned}
      \underset{p, q, s, r}{\mathrm{minimize}} \quad & \sum_{j = 1}^M\sum_{k: T_k \le T^j} \Ljk(s_j[k])\\
      \mathrm{subject\; to}
      \quad & \sum_{j \in \A_i}\sum_{k: T_k \le T_i} r_{ij}[k] \ge C_i\\
      \quad & \sum_{i \in \B_j \cap \Tk} r_{ij}[k] = s_j[k], r_{ij}[k] \ge 0\\
      \quad & i \in \B_j, p_i < q_j[k] \implies r_{ij}[k] = 0\\
      \quad & q_j[k] = \underset{i \in B_j \cap \Tk}{\text{max}} p_i,  s_j[k] = \Wjk(q_j[k]), q_j[k] \ge 0, p_i \ge 0.\\
    \end{aligned}
  \end{equation}

  As written, the final two lines of constraints are intractable.
  However, the cost function is independent of these constraints, and
  omitting them completely results in Problem
  \eqref{eqn:problem_finite}.  We show that for solutions of Problem
  \eqref{eqn:problem_finite}, there necessarily exists variables
  $p, q$ satisfying these additional constraints, and therefore that
  they can be omitted without affecting optimality.

  Since \eqref{eqn:problem_finite} is convex with linear constraints,
  the first order Lagrangian conditions are necessary and sufficient.
  We consider multipliers $\rho_i, \mu_{jk}, \theta_{ijk}$ and
  Lagrangian
  \begin{align*}
    \L(s, r, \rho, \mu, \theta)
    &= \sum_{j = 1}^M\sum_{k: T_k \le T^j} \Ljk(s_j[k]) + \sum_{i = 1}^N \rho_i \Bigl(C_i - \sum_{j \in \A_i}\sum_{k: T_k \le T_i} r_{ij}[k]\Bigl)\\
    &\quad + \sum_{j = 1}^M\sum_{k:T_k \le T^j}\mu_{jk}\Bigl(\sum_{i \in \B_j \cap \Tk} r_{ij}[k] - s_j[k]\Bigr) - \sum_{i = 1}^N\sum_{j \in \A_i}\sum_{k:T_k \le T_i} \theta_{ijk}r_{ij}[k].
  \end{align*}

  Suppose that $s, r$ are optimal primal solutions -- we require
  multipliers satisfying $\rho_i \ge 0, \theta_{ijk} \ge 0,$
  complementary slackness, and the stationarity conditions
  $\frac{\partial\L}{\partial r_{ij}[k]} = \frac{\partial\L}{\partial
    s_j[k]} = 0$:
  \begin{align*}
    \mu_{jk} &= \theta_{ijk} + \rho_i\\
    \Ljk'(s_j[k]) &= \mu_{jk},
  \end{align*}

  where it is implicit that $(i, j, k)$ must satisfy
  $j \in \A_i, T_k \le T_i$.  From the first equation, it must be that
  $\mu_{jk} \ge 0$ and therefore we can solve the second equation to
  obtain $s_j[k] = \Wjk(\mu_{jk})$.  We will write the dual problem
  and deduce that
  $\mu_{jk} = \underset{i \in \B_j \cap \Tk}{\text{max}}\; \rho_i
  \defeq \rho_j^\star[k]$.  Substituting the above stationarity
  conditions into the Lagrangian we have the dual:
  \begin{equation}
    \label{eqn:dual_problem}
    \begin{aligned}
      \underset{\rho, \mu, \theta}{\mathrm{maximize}} \quad & \sum_{j = 1}^M\sum_{k: T_k \le T^j} \Bigl[\fjk(\mu_{jk}) - \mu_{jk}\Wjk(\mu_{jk})\Bigr] + \sum_{i = 1}^N \rho_iC_i\\
      \mathrm{subject\; to}
      \quad & \mu_{jk} = \theta_{ijk} + \rho_i\\
      \quad & \theta_{ijk} \ge 0, \rho_i \ge 0.\\
    \end{aligned}
    \tag{$D^\star$}
  \end{equation}

  The objective function has derivative w.r.t. $\mu_{jk}$ of simply
  $-\Wjk(\mu_{jk}) < 0$ and is therefore monotone decreasing.
  Assuming $\rho_i$ is optimal for \eqref{eqn:dual_problem}, the
  variable $\theta_{ijk}$ is simply a slack variable which requires
  $\mu_{jk} \ge \rho_i$ for $i, j, k$ satisfying
  $j \in \A_i, T_k \le T_i$.  Therefore, by the monotonicity of the
  objective, the optimal $\mu_{jk}$ must be the smallest feasible,
  which is
  $\mu_{jk} = \underset{i \in \B_j \cap \Tk}{\text{max}}\; \rho_i$.
  It follows that $\theta_{ijk} = \rho_j^\star[k] - \rho_i$.

  These dual variables necessarily satisfy the constraints of Problem
  \eqref{eqn:finite_problem_full} with $p_i = \rho_i \ge 0$,
  $q_j[k] = \rho_j^\star[k]$ since by complementary slackness and the
  form of $\theta_{ijk}$ above, we have
  $p_i < p_j^\star[k] \implies r_{ij}[k] = 0$.

  That the solution is globally optimal follows from the fact that
  solutions of \eqref{eqn:problem_finite}, transformed to continuous
  solutions as described, satisfy the conditions of Proposition
  \ref{prop:optimal_allocation} and are therefore globally optimal for
  \eqref{eqn:ct_problem_cvx} by Proposition
  \ref{prop:global_optimality}.  These solutions can be transformed
  into solutions of the equivalent problem \eqref{eqn:ct_problem} as
  seen in Section \ref{sec:convex_reformulation}.

\section{Simulation (Additional Details)}
In this section we provide additional details on the methods used to
produce the results of Section \ref{sec:simulations}.

\subsection{Estimating Supply Curves}
\label{sec:estimating_supply_curves}
\begin{figure}
  \centering
  \caption{Estimated Supply Curves and Costs}
  \label{fig:supply_curve_estimates}

 { \begin{subfigure}[b]{0.4\textwidth}
    \caption{Supply Rate $\lambda(t)$}
    \label{fig:supply_rate_estimate}
    \includegraphics[width=\textwidth]{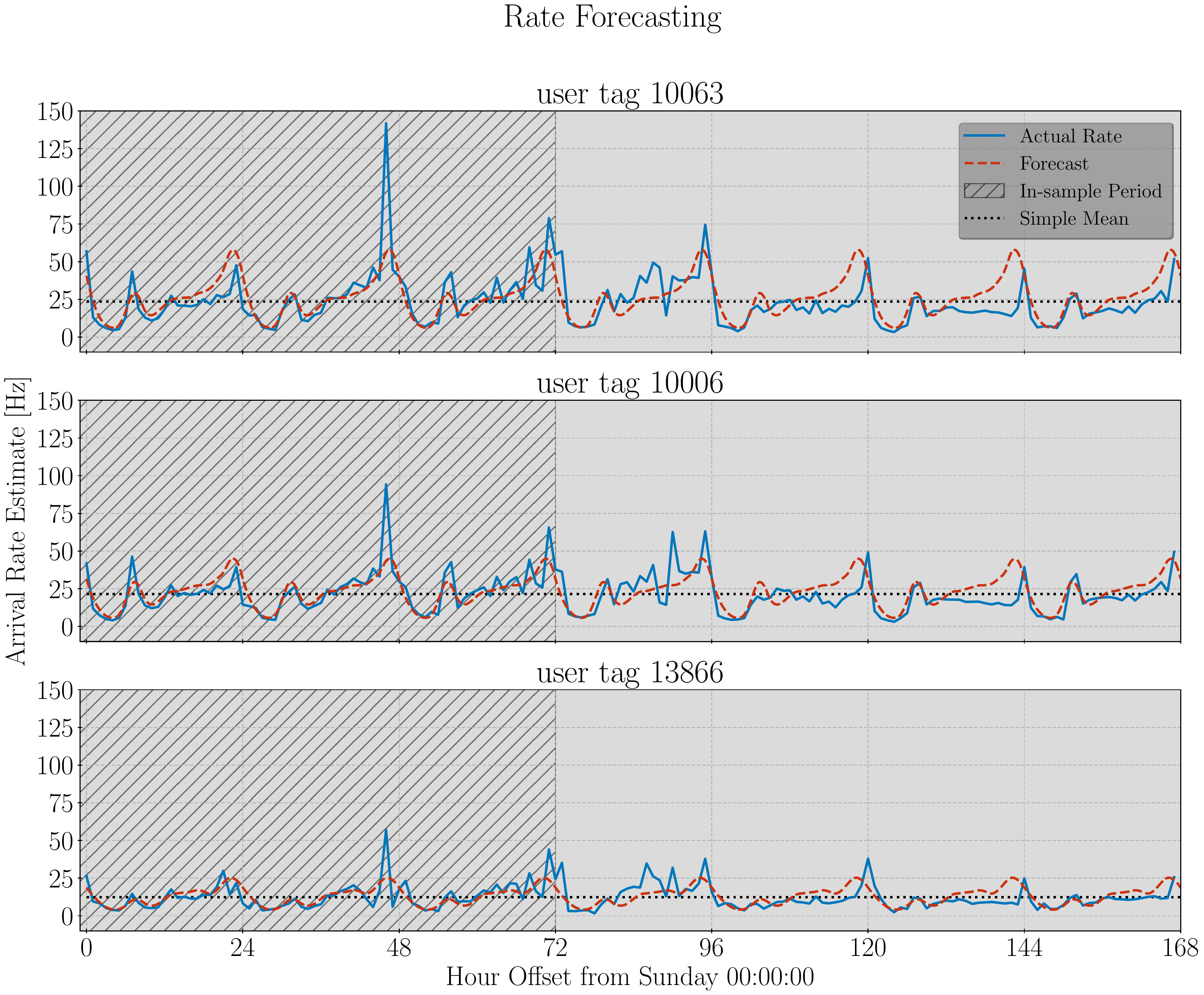}
  \end{subfigure}
  \begin{subfigure}[b]{0.4\textwidth}
    \caption{Win Probability Estimates}
    \label{fig:win_prob_estimate}
    \includegraphics[width=\textwidth]{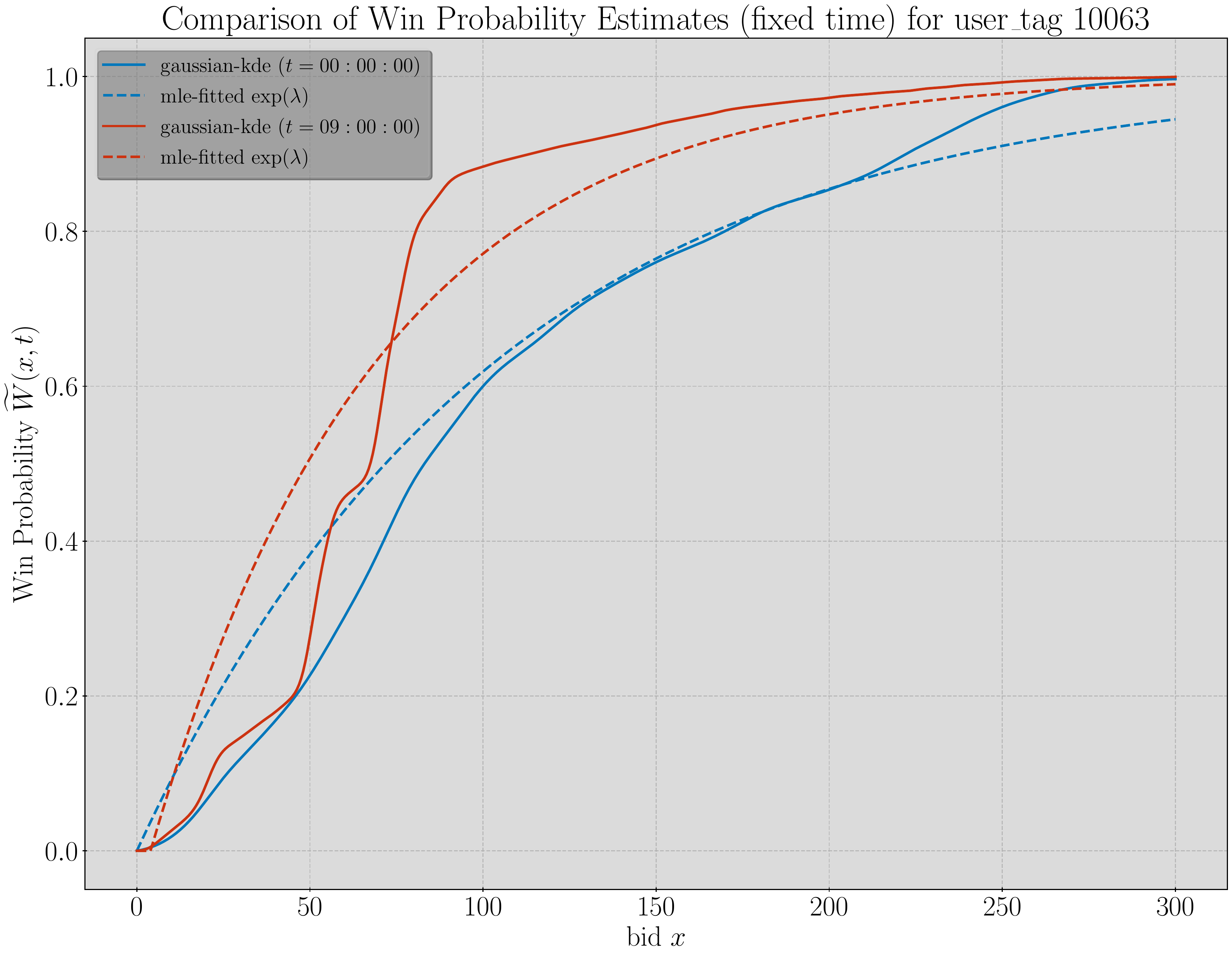}
  \end{subfigure}

  % \begin{subfigure}[b]{0.4\textwidth}
  %   \caption{Supply Curve Surface $W(x, t)$}
  %   \label{fig:supply_curve_surface}
  %   \includegraphics[width=\textwidth]{supply_curve_surface_kde.pdf}
  % \end{subfigure}
  % \begin{subfigure}[b]{0.4\textwidth}
  %   \caption{Cost Curve Surface $f(x, t)$}
  %   \label{fig:cost_curve_surface}
  %   \includegraphics[width=\textwidth]{cost_surface_kde.pdf}
  % \end{subfigure}

  \footnotesize{Cost and supply curves estimated from iPinYou data
    from the first 3 days of season 2.  $(a)$ Item arrival rates and
    the corresponding forecasts.  The hatched region indicates an
    in-sample period with the remainder being out-of-sample.  Our
    simulations run on similar 3 day periods with a 12 hour sliding
    window for a total of 9 periods of 72 hours each.  $(b)$
    Estimates of the win probability function $\widetilde{W}(x, t)$
    for $t = 00:00:00$ (blue) and $t = 09:00:00$ (red).  We compare
    Gaussian KDE (solid line) with a parametric Exponential CDF
    (dashed line).}}
    
    \end{figure}

The iPinYou dataset consists of impression data derived from a real
DSP and includes information about bidding prices, market prices, and
user characteristics.  We focus on the \textit{season two} data (a
week long period $2013$-$06$-$06$ to $2013$-$06$-$12$).  In all cases,
our supply curve estimates are $24$h-periodic in time and therefore
account for the natural daily (but not weekly) cycles in prices and
arrival rates.  Since this paper does not focus on the estimation of
supply curves, we apply a simple estimation procedure using the entire
dataset as input.  Though this has the effect of leaking some
information from the future, the estimation of supply curves is not
subject to optimization, limiting the impact of this leakage.
Moreover, the dataset is averaged into a single $24$h periodic
function and extended through periodicity.  It is reasonable to
believe that the previous week's (out of sample) data would provide
similar results.  Figure \ref{fig:supply_curve_estimates} provides an
illustration of estimated supply curves where only $72$h of data is
used to forecast the remaining $96$h for purposes of illustration.

Provided with the iPinYou dataset is a \texttt{user\_tag} which,
according to \cite{zhang2014real} is ``[a segment] in iPinYou’s
proprietary audience database''.  We therefore use the
\texttt{user\_tag} property as the ``item types'', focusing on the
five most common tags: \texttt{10063}, \texttt{10006}, \texttt{13866},
\texttt{10024}, and \texttt{10083}.

\subsection{Estimating the Supply Rate}
In order to estimate the supply rate $\lambda_j(t)$ for each user tag,
we have taken the inverse of the average of the time differences
between arrival instants in each hour of the day, after removing
outliers.  Calculating the number of arrivals over an hour long period
is not adequate as there appear to be large consistent gaps in arrival
times: we suspect that the dataset was subsampled prior to being
released.

This calculation results in estimates
$\lambda_j[0], ..., \lambda_j[23]$ with time denoted in hours.  The continuous
estimate was subsequently formed by smoothly interpolating between
these points with a $24-$periodic boundary, resulting in a function
$\tilde{\lambda}_j(t)$ defined on $[0, 24]$.  A forecast for the average
supply rate at time $t$ is obtained via
$\lambda_j(t) = \tilde{\lambda}(t\; \text{mod}\; 24)$.

An illustrative example for tags \texttt{10063}, \texttt{10006}, and
\texttt{13866} is provided in Figure \ref{fig:supply_rate_estimate}.

\subsection{Estimating Win Probabilities}
Similarly to the supply rate estimates, we estimate an average win
probability function for each $t \in {0, \ldots, 23}$ and then
smoothly interpolate along $t$ to estimate a $24-$periodic function
$\widetilde{W}_j(x, t)$ indicating the probability of winning an
impression of type (user tag) $j$ arriving at time $t$ given a bid
$x$.

The estimate of $x \mapsto \widetilde{W}_j(x, t)$ is obtained by
smoothing the histogram with a Gaussian kernel (bandwidth chosen
simply by the Normal Reference Rule
\cite[Chap.~6.3]{wasserman2006all}) for each \texttt{market\_price}
data point falling into the hour long window.  The results of this
procedure, as well as a comparison to a parametric estimate with an
Exponential density are given in Figure \ref{fig:win_prob_estimate}.

The \texttt{market\_price} attribute in the dataset corresponds to the
price actually paid in the second price auction.  We have not
accounted for the affects of censoring -- since the DSP collected the
dataset with large bids intended to win most impressions that were bid
on, this isn't a significant factor.

\subsection{Cost and Supply Curves}
The supply curve $W_j(x, t)$ is simply the product of the supply rate
$\lambda_j(t)$ and the win probability $\widetilde{W}_j(x, t)$.  The
cost curve $f(x, t) \defeq \int_0^x uW'(u, t)\d u$ is derived from the
supply function where we have used numerical integration and
differentiation to estimate $f$ on a grid and subsequently extended
$f$ to the entire surface via interpolation.

\subsection{Simulating the Bidding Process}
\label{sec:bidding_process_simulation}

The simulations of Section \ref{sec:simulations} are obtained by
storing the hour-by-hour inter-arrival and price data for each item
type $j \in [M]$ and sampling uniformly from these datasets.  At
simulation time $t \in \R_+$ we sample an inter-arrival time
$\Delta t$ and price $P$ from the data for hour
$\lfloor t \rfloor + 1$ with probability $t - \lfloor t \rfloor$ and
otherwise from the data for hour $\lfloor t \rfloor$.  A bid is
solicited from a bidder (an implementation of
\eqref{eqn:problem_finite} or the algorithm of
\cite{marbach_bidding_2020}) and if the bid exceeds $P$ the bidder
allocates that item to the fulfillment of a contract.  The simulation
time is them updated to $t + \Delta t$ and the process continues.

% \newpage

% This can be done through the package's \caption
% \vspace{12pt}
% \begin{center}
%     {\bf Algorithm}: Bidding Simulation
% \end{center}
% \vspace{-6pt}

\begin{algorithm}
  \SetKwInOut{Input}{input}
  \SetKwInOut{Output}{output}
  \SetKwInOut{Initialize}{initialize}
  \DontPrintSemicolon

  \SetKwFunction{FSample}{Sample-Dataset}

  \BlankLine
  \caption{Bidding Simulation}
  \label{alg:bidding_process}
  
  \Input{A $\text{Bidder}$ derived from Section \ref{sec:general_case} and solution to Problem \eqref{eqn:problem_finite}}
  \Output{Recording of Bidder's item allocations to process into normalized acquisition curves $\tilde{c}(t)$}

  \texttt{// Initialize:}\;
  $Q \leftarrow \text{Priority-Queue}([\ ])$ \texttt{  // Sort by time}\;
  $t \leftarrow 0$ \texttt{  // The ``current'' time}\;
  \For{$j \in [M]$} {
    \texttt{  // Sample an interarrival time and a price}\;
    $(\Delta t, P) \leftarrow\ $\texttt{Sample-Dataset($t$,$j$)}\;
    $Q$\texttt{.push($(t + \Delta t, P, j)$)}\;
  }

  \BlankLine
  \texttt{// Simulate bidding process:}\;
  \While{$t < T_{\text{end}}$} {
    $t, P, j \leftarrow Q$\texttt{.pop()}\;
    $b \leftarrow \text{Bidder}$\texttt{.solicit\_bid($t$,$j$)} \texttt{  // Ask for a bid on type $j$ at time $t$}\;
    \If{$b \ge P$} {
      $\text{Bidder}$\texttt{.award\_item($t$,$j$)} \texttt{  // Allocate items for winning bids}\;
    }
    $(\Delta t, P) \leftarrow\ $\texttt{Sample-Dataset($t$,$j$)} \texttt{  // Append next $(t, P)$ pair for $j$ to $Q$}\;
      $Q$\texttt{.push($(t + \Delta t, P, j)$)}\;
  }

  \BlankLine
  \SetKwProg{Fn}{Function}{\texttt{($t$,$j$):}}{\KwRet}
  \Fn{\FSample}{
    $p \leftarrow t - \lfloor t \rfloor$\;
    $U \sim \mathcal{U}(0, 1)$  \texttt{  // Interpolate between hours}\;
    \If{$p \le U$}{
      $h \leftarrow \lfloor t \rfloor$\;
    }
    \Else{
      $ h \leftarrow \lfloor t \rfloor + 1$\;
    }
    $\Delta t \leftarrow $\texttt{ Sample-Interarrivals(hour=$h$,type=$j$)}\;
    $P \leftarrow $\texttt{ Sample-Prices(hour=$h$,type=$j$)}\;
    \Return $(\Delta t, P)$\;

  }
\end{algorithm}

\end{document}